\documentclass[%
 reprint,
floatfix,
superscriptaddress,
showpacs,preprintnumbers,
 amsmath,amssymb,
 aps,
pra,
]{revtex4-2}
\DeclareUnicodeCharacter{2212}{-}
\usepackage{xcolor}
\usepackage{times}
\usepackage{amssymb}
\usepackage{dsfont}
\usepackage{tikz}
\usepackage{graphicx}% Include figure files
\usepackage{dcolumn}% Align table columns on decimal point
\usepackage{bm}% bold math
\usepackage{epstopdf}
\usepackage[colorlinks]{hyperref}% add hypertext capabilities
\hypersetup{colorlinks = true,
linkcolor = blue,
filecolor = blue,
urlcolor = blue,
citecolor = blue}
\newcommand*\chem[1]{\ensuremath{\mathrm{#1}}}

\begin{document}
\title{Spontaneous layer selective Mott phase in the bilayer Hubbard model}
\author{Emile Pangburn}
\affiliation{Universit\'{e} Paris-Saclay, Institut de Physique Th\'eorique,  CEA, CNRS, F-91191 Gif-sur-Yvette, France}
\author{Louis Haurie}
\affiliation{Universit\'{e} Paris-Saclay, Institut de Physique Th\'eorique,  CEA, CNRS, F-91191 Gif-sur-Yvette, France}

\author{Sébastien Burdin}
\affiliation{Université de Bordeaux, CNRS, LOMA, UMR 5798, F-33400 Talence, France}

\author{Catherine P\'epin}
\affiliation{Universit\'{e} Paris-Saclay, Institut de Physique Th\'eorique,  CEA, CNRS, F-91191 Gif-sur-Yvette, France}

\author{Anurag Banerjee}
\affiliation{Universit\'{e} Paris-Saclay, Institut de Physique Th\'eorique,  CEA, CNRS, F-91191 Gif-sur-Yvette, France}
% Please give the surname of the lead author for the running footer
\begin{abstract}
Quantum materials featuring both itinerant and localized degrees of freedom exhibit numerous exotic phases and transitions that deviate from the Ginzburg-Landau paradigm. This work uses the composite operator formalism to examine the bilayer strongly correlated Hubbard model. We observe the spontaneous breaking of layer symmetry, where the electron density in one of the layer reaches half-filling, resulting in a layer selective Mott phase (LSMP). This broken symmetry phase becomes unstable at a critical average electronic density away from half-filling. Furthermore, significant layer differentiation persists up to a moderate inter-layer hopping, beyond which the system abruptly transitions to an  layer uniform phase (LUP). In the LSMP phase, the electrons in the two layers are weakly hybridized, resulting in a small Fermi surface. The volume of the Fermi surface jumps at the transition from the LSMP to the uniform phase. We also discuss the physical mechanisms leading to the collapse of the LSMP phase under different perturbations.
\end{abstract}
\maketitle
\section{Introduction}
\label{introduction}
Investigating strong electronic correlations is a central challenge in quantum matter, underlying many nontrivial phenomena, such as high-$T_c$ superconductivity~\cite{bednorz1986possible,DagottoRMP,Keimer2015}, Mott insulator~\cite{mott1968metal,Cai2016,LucaMottPRL,MottVO2}, and heavy Fermions~\cite{HFLOld0,HFSC,Custers2003,Schröder2000}, among others. The significant interactions between different degrees of freedom of quantum materials lead to multiple broken symmetry phases, resulting in an extremely rich phase diagram. 

Orbital degrees of freedom can lead to intriguing phenomena in strongly correlated quantum matter such as spontaneous orbital symmetry breaking, Hund metallicity~\cite{GeorgesHund24}. The orbital selective Mott phase (OMSP) was first proposed to explain the coexistence of metallic and magnetic properties in ruthenates~\cite{anisimov2002orbital,Ruth1}  and has since been recognized as key for understanding the normal~\cite{si2016high}  and superconducting properties~\cite{kreisel2017orbital,sprau2017discovery}  of iron-based superconductors. It is also crucial for understanding the local versus nonlocal duality in $f$-electron-based compounds~\cite{zwicknagl2003dual}. The OSMP is characterized by the selective localization of electrons in specific orbitals, while electrons in other orbitals remain itinerant. Initial evidence came from Angle-resolved photoemission spectroscopy (ARPES) studies~\cite{neupane2009observation} in \chem{Ca_{1.8}Sr_{0.2}RuO_4}, followed by similar observations in iron-chalcogenides~\cite{yi2015observation}. In these materials, the spectral weight of the $d_{xy}$ orbitals vanishes upon cooling, while the spectral weight of other $3d$ orbitals remains non-zero. More recently, theoretical proposals suggest that OSMPlike phase could also emerge in twisted bilayer systems such as twisted dichalcogenides~\cite{dalal2021orbitally} and twisted trilayer graphene~\cite{ramires2021emulating}.

The theoretical model often used to describe this phase is a multiorbital Hubbard model with intraorbital and interorbital on-site repulsion $U$ and $U^\prime$, and Hund's coupling $J_H$ \cite{georges2013strong}. In the strong correlation regime, $U$ prevents double occupancy, favoring antiferromagnetism on a square lattice, while $J_H$ promotes local spin alignment across different orbitals, corresponding to ferromagnetic interaction. Both $U^\prime$ and $J_H$ are correlated interactions between orbitals. Previous works \cite{de2011hund,de2011janus} highlight Hund's coupling as the primary factor in stabilizing the OSMP phase in non-degenerate correlated orbitals by suppressing orbital fluctuations through spin locking.

The orbital selective phase can be defined as the complete localization of one orbital entirely decoupled from the other metallic orbitals~\cite{kugler2022orbital}. The stability of the OSMP phase with interorbital hoppings remains a subject of debate. While there is agreement among various methods that an OSMP exists when the hopping between different orbitals is negligible, the situation becomes more intricate when it is non-zero. The slave-spin formalism indicates the stability of the OSMP phase with inter-orbital hopping~\cite{komijani2017analytical,yu2017orbital}, with the hybridization of the localized orbitals to the others renormalized to zero. Single-site dynamical mean field theory (DMFT) claims~\cite{kugler2022orbital} that any inter-orbital hopping should lead to a finite hybridization. Nevertheless, both methods agree that even with finite hybridization, orbital differentiation exists with a significant difference in effective mass. 

This paper investigates the degenerate bilayer Hubbard model \cite{beach2011orbital,kancharla2007band,golor2014ground,lee2014competition,goldberger2024dynamical}, with only on-site repulsion $U$.  Since we do not consider Hund's coupling necessary for stabilizing the OSMP phase~\cite{de2011hund,de2011janus}, we refer to the resulting symmetry broken phase as layer selective Mott phase (LSMP) to distinguish it from the OSMP. Such layer-selective Mott phases can potentially be important for delafossites~\cite{Lechermann_def} and nickelates~\cite{Lechermann_nick} and other layered systems. We examine the effect of finite inter-layer hopping while varying the total electron density in the strong correlation regime. In this regime, it is suitable to start from the atomic Hamiltonian $(t=0)$ and generate a set of quasiparticle operators for the ground state. In these schemes, electrons in a strongly interacting regime are generally considered a sum of weakly interacting excitations~\cite{ovchinnikov2004hubbard}. These excitations correspond to holons and doublons in the strongly correlated limit of the one-band Hubbard model, both of which have identical quantum numbers to the electron. This results in the emergence of the lower and upper Hubbard bands, characterized predominantly by holons and doublons, respectively. Such schemes can be contrasted with slave-boson methods~\cite{kotliar1986new,wen1996theory,florens2004slave} where the electron fractionalizes into emergent operators that carry fractions of the electron's charge and spin. This introduces gauge freedom, making the model equivalent to solving gauge theories.

Using the composite operator method (COM) \cite{Stanescu0, hubbard1963, Beenen0, avella2011composite,avella2014composite,Haurie_2024}, we explore the strongly correlated limit of the two layer Hubbard model. This technique has previously been applied to specific solutions of the two-band Hubbard model \cite{avella2007two}, demonstrating the existence of a selective Mott phase. The objective of this paper is to perform a detailed study of the physical properties and layer symmetry-breaking phase in such a model. Our central finding is the emergence of a spontaneous layer differentiation phase for moderate values of interlayer hopping without explicit symmetry-breaking parameters between the two layers \cite{beach2011orbital}. Our results offer a complementary perspective to slave-boson and DMFT approaches in similar models and provide novel insights into the impact of finite inter-layer hopping on the LSMP phase and its abrupt destabilization. Additionally, we investigate the correspondence between the Kondo lattice model and bilayer Hubbard model~\cite{pepin2007kondo, pepin2008selective}, particularly examining whether complete Kondo breakdown \cite{coleman2001fermi} occurs in the LSMP phase.

The plan for the rest of the paper is as follows. The next section discusses the model and the composite operator formalism for the two-band Hubbard model. The detailed expressions for self-consistency equations and physical observables are presented in Appendix~(\ref{Appendix:Physical_Quantities}). We present the overall phase diagram for the strength of fixed interactions in Sec.~(\ref{sec:PhaseDiagramU20}). The subsequent subsections discuss the bands, the Fermi surface, the density of states, and quasiparticle residues identifying the two phases. We conclude with a discussion of the results in Sec.~(\ref{sec:Discussions}).

\section{Model and method\label{sec:Model_Method}}
We work with the two-band Hubbard model 
\begin{align}
\mathcal{H}=&-\sum \limits_{ij\alpha\sigma} t_{ij}^\alpha c^\dagger_{i\alpha\sigma} c_{j\alpha\sigma} -\lambda \sum \limits_{i \sigma} (c^\dagger_{i x \sigma} c_{i y \sigma} + c^\dagger_{i y \sigma} c_{i x \sigma}) &\\
\nonumber& +  \sum \limits_{i \alpha} U^\alpha \hat{n}_{i \alpha \uparrow} \hat{n}_{i \alpha \downarrow}- \mu\sum \limits_{i \sigma \alpha}  \hat{n}_{i \alpha \sigma} &
\end{align}
Here, $c_{i\alpha\sigma}$ $(c^\dagger_{i\alpha\sigma})$ annihilates (creates) an electron at site $i$ with layer $\alpha$ and spin $\sigma$, where $\sigma=\uparrow,\downarrow$ for spin-$1/2$ electrons and ${\alpha=x,y}$ in a square lattice. The number operator is defined as $\hat{n}_{i\alpha\sigma}=c^\dagger_{i\alpha\sigma}c_{i\alpha\sigma}$, and the local electron density is the expectation value of the number operator,
$n_{i\alpha\sigma} = \langle\hat{n}_{i\alpha\sigma}\rangle$.

The first term represents intra-layer hopping between neighboring sites $i$ and $j$, set to $t_{ij}=t$ if $i$ and $j$ are nearest neighbors; otherwise, zero. We set $t^x=t^y=t$ in this work, and all our energy scales are in units of $t=1$. The second term describes on-site inter-layer hopping with a coupling constant $\lambda$. The third term accounts for intra-layer on-site repulsion between electrons. We assume that this intra-layer repulsion has an equal strength for both layers $U^x=U^y=U>0$. The chemical potential $\mu$ fixes the total electron density of the system $n=(1/N)\sum_{i,\alpha,\sigma} \langle \hat{n}_{i\alpha\sigma} \rangle$ which can vary from $n \in [0,4]$. Here is the number of lattice points in the system.
Our analysis of the system is valid in the strong correlation regime where $U\gg t, \lambda$.

\subsection{Composite operators - Equation of Motion}
In the strong correlation regime, the excitation of the singly occupied ground state are given by holon ($\xi$) and doublon ($\eta$) single layer operators
\begin{align}
\label{eq:Xi_Operator}&\xi_{i,\alpha\sigma}=c_{i\alpha\sigma} (1-n_{i,\alpha\overline{\sigma}}), \\ 
\label{eq:Eta_Operator}&\eta_{i,\alpha\sigma}=c_{i\alpha\sigma}n_{i,\alpha\overline{\sigma}}.
\end{align}

Since we are focusing on an interacting model with only on-site, intra-layer interactions, the $\xi$ and $\eta$ operators represent the most relevant excitations near half-filling. To capture the physics of more complex models, such as those including significant inter-layer Coulomb interactions, the operator basis must be expanded. This is the case in multi-orbital models with a Hund term, where additional bands can be observed~\cite{stadler2019hundness,sroda2023hund}. In contrast, our approximation will always yield four bands for two layers.
The composite operator method (COM) \cite{avella2011composite} is an equation of motion technique to compute the electronic two-point Green's function $\mathcal{G}(\mathbf{k},\omega)$ and hence other physical quantities. For the two-band Hubbard model with only intra-layer electronic repulsion, the paramagnetic composite operator basis is defined by
\begin{align}
\mathbf{\Psi}=\left(... ,\xi_{ix\sigma},\eta_{ix\sigma},\xi_{iy\sigma},\eta_{iy\sigma}, ... \right)^T,
\label{Eq:CompOP}
\end{align}
leading to a $4N$ component vector.
The matrix form of the Green's function in the imaginary time $\tau$ is written as
\begin{align}
    \mathds{G}(\tau)=- \left\langle T_\tau \left( \mathbf{\Psi}( \tau )  \mathbf{\Psi}^\dagger(0) \right)  \right\rangle,
\end{align}
Here $T_\tau$ is the time ordering operator. The equation of motion for the Green's function is
\begin{align}
    -\partial_\tau \mathds{G}(\tau) = &\delta(\tau)  \left\langle \{   \mathbf{\Psi}(0),\mathbf{\Psi}^\dagger(0) \}  \right\rangle -  \left\langle T_\tau \left( \mathbf{j}( \tau )  \mathbf{\Psi}^\dagger(0) \right)  \right\rangle,  \label{Eq:EOM_GIj}
%    -\partial_\tau \mathds{G}(\tau) &=-\delta(\tau) \mathds{I}(0) + \Theta(\tau) \mathds{M}(\tau)
%    \label{Eq:EOM_GIM}
\end{align}
where we defined the current as
${\mathbf{j}(\tau)=\left[ \mathcal{H} , \mathbf{\Psi}  \right]  (\tau)}$. Next we define the normalization matrix $\mathds{I}$ and $\mathds{M}$-matrix as
\begin{align}
    \mathds{I} &= \left\langle \{ \mathbf{\Psi}(0),\mathbf{\Psi}^\dagger(0) \}  \right\rangle,    \label{Eq:EOM_IMat}\\
    \mathds{M}(\tau) &=\left\langle T_\tau \left( \mathbf{j}(\tau) \mathbf{\Psi}^\dagger(0) \right) \right\rangle \label{Eq:EOM_MMat}.
\end{align}

To close the equations in this order, we focus solely on the current along the composite operator, as defined in Eq. (\ref{Eq:CompOP}), we obtain
\begin{align}
\mathbf{j}(\tau)\approx\mathds{E} \mathbf{\Psi}(\tau).
\label{Eq:Apprxj}
\end{align}
Here, $\mathds{E}$ represents the energy matrix, ensuring the current remains proportional and aligned with the defined basis.
Using Eq.~(\ref{Eq:EOM_GIj}), Eq.~(\ref{Eq:EOM_IMat}), Eq.~(\ref{Eq:EOM_MMat}) and Eq.~(\ref{Eq:Apprxj}) we get the definition of the $\mathds{E}$-matrix and $\mathds{M}(0)$-matrix
\begin{align}
    \mathds{M}(0)&= \left\langle \{ \mathbf{j}(0),\mathbf{\Psi}^\dagger(0) \}  \right\rangle,  \label{Eq:M0}    \\
    \mathds{E}(0)&= \mathds{M}(0) \mathds{I}^{-1}.
    \label{Eq:E0}
\end{align}
Putting Eq.~(\ref{Eq:Apprxj}) in Eq.~(\ref{Eq:EOM_GIj}), we obtain
\begin{align}
   -\partial_\tau \mathds{G}(\tau) &= \delta(\tau)  \mathds{I} -  \mathds{E}(0) \left\langle T_\tau \left( \mathbf{\Psi}( \tau )  \mathbf{\Psi}^\dagger(0) \right)  \right\rangle.
%    \partial_\tau \mathds{G}(\tau) &= -\delta(\tau)  \mathds{I} + \mathds{E}(0)\mathds{G}(\tau)
\end{align}
Next we Fourier transform to Matsubara frequency and after making appropriate analytic continuation, we get advanced and retarded Green's functions
\begin{align}
    \mathds{G}^{R/A}(\omega)=\left[ (\omega \pm i\epsilon) \mathds{1} - \mathds{E} \right]^{-1} \mathds{I},
    \label{Eq:Grealomega}
\end{align}
where $\mathds{1}$ is the identity matrix, and $\epsilon>0$ is small. Now we need to calculate the $\mathds{M}$, $\mathds{I}$ and hence $\mathds{E}$-matrix to obtain the composite Green's function. These matrices are $4N\times4N$ matrices that incorporate the lattice structure. In the following, we exploit $C_4$ rotational symmetry to reduce the number of self-consistent parameters and translation invariance to perform a Fourier transform and conduct the calculations in momentum space. Consequently, these matrices become block diagonal in the momentum representation, similar to the retarded and advanced Green's functions 
\begin{align}
    \mathds{G}^{R/A}(\mathbf{k},\omega)=\left[(\omega \pm i\epsilon) \mathds{1} - \mathds{E}(\mathbf{k})  \right]^{-1} \mathds{I}.
    \label{Eq:Gkomega}
\end{align}
The poles of $\mathds{G}^{R/A}(\mathbf{k},\omega)$ correspond to the eigenvalues $E^i(\mathbf{k})$ of $\mathds{E}(\mathbf{k})$.
 We calculate the matrices for the two-band Hubbard model in the next subsection.
\subsection{Computation of E and Matrix}
Using the approximation for computing the current Eq.(\ref{Eq:Apprxj}) along with the definition of the M-matrix Eq.~(\ref{Eq:M0})  we arrive at the components of the M-matrix. 
\begin{align}
\mathds{M}_{11}(\mathbf{k})&=-\mu(1-n_x/2)-4te_{xx}-\lambda e_{xy}&\\[2pt]
\nonumber&-\gamma(\mathbf{k})(1-n_x+p_{xx})=\mathds{M}^\prime_{33}(\mathbf{k})&\\[5pt]
\mathds{M}_{12}(\mathbf{k})&=4te_{xx}+\lambda e_{xy}-\gamma(\mathbf{k})(n_x/2-p_{xx})&\\[2pt]
\nonumber&=\mathds{M}_{21}(\mathbf{k})=\mathds{M}^\prime_{34}(\mathbf{k})=\mathds{M}^\prime_{43}(\mathbf{k})&\\[5pt]
\mathds{M}_{22}(\mathbf{k})&=-(\mu-U)n_x/2-4te_{xx}-\lambda e_{xy}&\\[2pt]
\nonumber&-\gamma(\mathbf{k})(1-n_x+p_{xx})=\mathds{M}^\prime_{44}(\mathbf{k})&\\[5pt]
%\mathds{M}_{33}(\mathbf{k})&=-\mu(1-n_y/2)-4te_{yy}-\lambda e_{yx}&\\[2pt]
%\nonumber&-\gamma(\mathbf{k})(1-n_y+p_{yy})&\\[5pt]
%\mathds{M}_{34}(\mathbf{k})&=4te_{yy}+\lambda e_{yx}&\\[2pt]
%\nonumber&-\gamma(\mathbf{k})(n_y/2-p_{yy})=\mathds{M}_{43}(\mathbf{k})&\\[5pt]
%\mathds{M}_{44}(\mathbf{k})&=-(\mu-U)n_y/2-4te_{yy}-\lambda e_{yx}&\\[2pt]
%\nonumber&-\gamma(\mathbf{k})(1-n_y+p_{yy})&\\[5pt]
\mathds{M}_{13}(\mathbf{k})&=-\lambda\left(1-n_x/2-n_y/2+p_{xy}\right)=\mathds{M}_{31}(\mathbf{k})&\\[5pt]
\mathds{M}_{23}(\mathbf{k})&=-\lambda\left(n_x/2-p_{xy}\right) \nonumber \\&=\mathds{M}_{32}(\mathbf{k})=\mathds{M}^\prime_{14}(\mathbf{k})=\mathds{M}^\prime_{41}(\mathbf{k})&\\[5pt]
%\mathds{M}_{14}(\mathbf{k})&=-\lambda\left(n_y/2-p_{xy}\right)=\mathds{M}_{41}(\mathbf{k})&\\[5pt]
\mathds{M}_{24}(\mathbf{k})&=-\lambda p_{xy}=\mathds{M}_{42}(\mathbf{k})&
\end{align}
Here $\gamma(\mathbf{k})=2t \left( \cos(k_x) + \cos(k_y) \right)$. Moreover, in $\mathds{M}^\prime$, the layer indices $x$ and $y$ are interchanged compared to $\mathds{M}$. For instance, $\mathds{M}_{33}$ has the same expression as $\mathds{M}_{11}$ with all the subscript $x$ and $y$ interchanged. We also define average layer electron density as
\begin{align}
    n_\alpha=\frac{1}{N}\sum_{i,\sigma} \langle \hat{n}_{i,\alpha,\sigma} \rangle. 
\end{align}
Furthermore, we introduced the following bond-expectation value averaged over nearest-neighbor. 
\begin{align}
e_{\alpha \beta}=\frac{1}{4N}\sum_{\langle i,j \rangle}  &\left( \langle\xi_{j \alpha \sigma}\xi^\dagger_{i \beta\sigma}\rangle -\langle\eta_{j \alpha \sigma}\eta^\dagger_{i \beta \sigma}\rangle \right. \nonumber \\  &\left.+\langle \eta_{j\beta\sigma} \xi^\dagger_{i\alpha\sigma} \rangle - \langle \eta_{j\alpha \sigma}\xi^\dagger_{i\beta\sigma} \rangle \right)\\
p_{\alpha  \beta}=\frac{1}{4N}&\sum_{\langle i,j \rangle}\langle \hat{n}_{i \alpha \sigma}\hat{n}_{j \beta \sigma}\rangle+\langle S^-_{i \alpha }S^+_{j \beta}\rangle-\langle \Delta_{i \alpha }\Delta^\dagger_{j \beta}\rangle \label{eq:pab}
\end{align}
with $S_{i\alpha}^-=c^\dagger_{i\alpha\downarrow}c_{i\alpha\uparrow}$, $S_{i\alpha}^+=c^\dagger_{i\alpha\uparrow}c_{i\alpha\downarrow}$ and $\Delta_{i\alpha}=c_{i\alpha\uparrow}c_{i\alpha\downarrow}$.

The normalization matrix is diagonal and is given by
\begin{align}
\mathds{I}=\text{diag}\Big[\big(1-n_x/2,n_x/2,1-n_y/2,n_y/2\big)\Big]
\end{align}
\subsection{Correlation function and self-consistency}
The $M$-matrix and the $I$-Matrix depends on nine initially unknown parameters ($e_{\alpha \beta },p_{\alpha \beta},n_\alpha,\mu$). Note that to evaluate the unknown parameter $n_\alpha$ and $e_{\alpha \beta}$ we only need the single particle on-site and nearest neighbor inter-site correlation functions. 
\begin{align}
C^{ab}_{\vert \mathbf{r}_{i}-\mathbf{r}_j \vert}=-\frac{1}{2 \pi N i} \sum_{\mathbf{k}} &e^{i\mathbf{k}\cdot \left(\mathbf{r}_i-\mathbf{r}_j\right)} \int d\omega \left[ \big(1-n_F(\omega) \big)\nonumber \right.
\\ 
 & \left. \times \left(G^R(\mathbf{k},\omega)-G^A(\mathbf{k,\omega})\right)_{ab} \right].
\label{Eq:Call}
\end{align}
Here $a$ and $b$ are integers between $\{1,4\}$ and $n_F(\omega)$ is the usual Fermi-Dirac distribution.
Assuming $C_4$ rotation symmetry we can compute the correlation by performing the $\omega$ integral in Eq.~(\ref{Eq:Call}) analytically~\cite{avella2011composite}. 

Using the on-site ($C^{ab}_0$) and nearest neighbor ($C^{ab}_1$) correlation functions we can write the layer density as
\begin{align}
    n_x= 2 \left(1-\sum_{i,j=1}^2 C_0^{ij} \right),\\
    n_y=2 \left( 1-\sum_{i,j=3}^4 C_0^{ij} \right).
\end{align}
We define the difference in layer density as ${\delta n=(n_x-n_y)/2}$ as a self-consistent parameter, and fix the total density $n=n_x+n_y$ using the chemical potential $\mu$. Similarly the $e$-parameters can be recalculated by
\begin{align}
    e_{xx}&= C^{11}_1-C^{22}_1, \\
    e_{yy}&= C^{33}_1-C^{44}_1, \\ 
    e_{xy}&= C_0^{31}+C_0^{41}-C_0^{23}-C_0^{24},\\ 
    e_{yx}&= C_0^{13}+C_0^{23}-C_0^{41}-C_0^{42}.
\end{align}

However, $p_{\alpha\beta}$ is a two-particle correlation function that cannot be evaluated directly from single-particle correlation functions. For the one-band Hubbard model, $p$ has also been computed by enforcing the Pauli principle $\langle \xi_i\eta_{i}^\dagger\rangle=0$~\cite{avella1998hubbard,avella2007underdoped}. In the two-band Hubbard model, such approach leaves the computation of interlayer $p_{xy}$ ambiguous. Therefore, we evaluate $p_{\alpha\beta}$ for $\alpha,\beta\in\{x,y\}$ using the Roth decoupling scheme~\cite{roth1969electron}, which we outline in Appendix~(\ref{Appendix:Roth}). The $p$-parameter depends on three distinct nearest neighbor correlation functions as seen in Eq.~(\ref{eq:pab}). The expression for density-density correlations in terms of $C^{ab}_0$ and $C^{ab}_{1}$ is given in Eq.~(\ref{eq:nn}). Similarly the expression for $\langle S^-_{i \alpha }S^+_{j \beta}\rangle$ in terms of correlation functions is presented in Eq.~(\ref{eq:ss}) and $\langle \Delta_{i \alpha }\Delta^\dagger_{j \beta} \rangle$ is presented in Eq.~(\ref{eq:dd}). Using these one can recalculate two intra-layer and one inter-layer $p_{\alpha\beta}$.

Subsequently, we solve these nine coupled self-consistency equations. In the layer-selective Mott phase, one of the layers becomes half-filled. Therefore, we initialize with the guess $\delta n = (2-n)$ and solve the self-consistency equations. If the converged solution has one of the layer densities approaching half-filling $n_\alpha \rightarrow 1$, then that solution represents the LSMP phase. In the literature \cite{de2005orbital,yu2013orbital,kugler2022orbital}, the LSMP phase is defined by the vanishing hybridization between the localized electrons in a given layer and the other itinerant layers. If the localized layer has a density very close to, but not exactly, $1$ due to finite hybridization, then the system's physical properties are effectively identical to those of a conventional LSMP phase for all practical purposes. In the following, we will extend the term LSMP to include such phases. Conversely, a solution with $\delta n=0$ represents a uniform phase (LUP). We compare the energy per site $E_s$ of these two solutions to identify the lower energy solution.

\section{Results}
\label{sec:Res}
We perform the calculations for a square lattice with linear dimension $L=300$ with periodic boundary conditions. The repulsion strength is fixed at $U=20 t$ for the results presented in the main text. However, we have ensured the reliability of our qualitative conclusions across various repulsion strengths ranging from $U=9t$ to $U=20t$. Our investigation spans a range of average electronic densities between $n=1.50$ and $n=1.94$. We focus on discerning layer symmetry breaking within systems invariant under spin-rotation and translation.

\subsection{Phase diagram}
\label{sec:PhaseDiagramU20}
\begin{figure}[ht]
\includegraphics[width=0.499\textwidth]{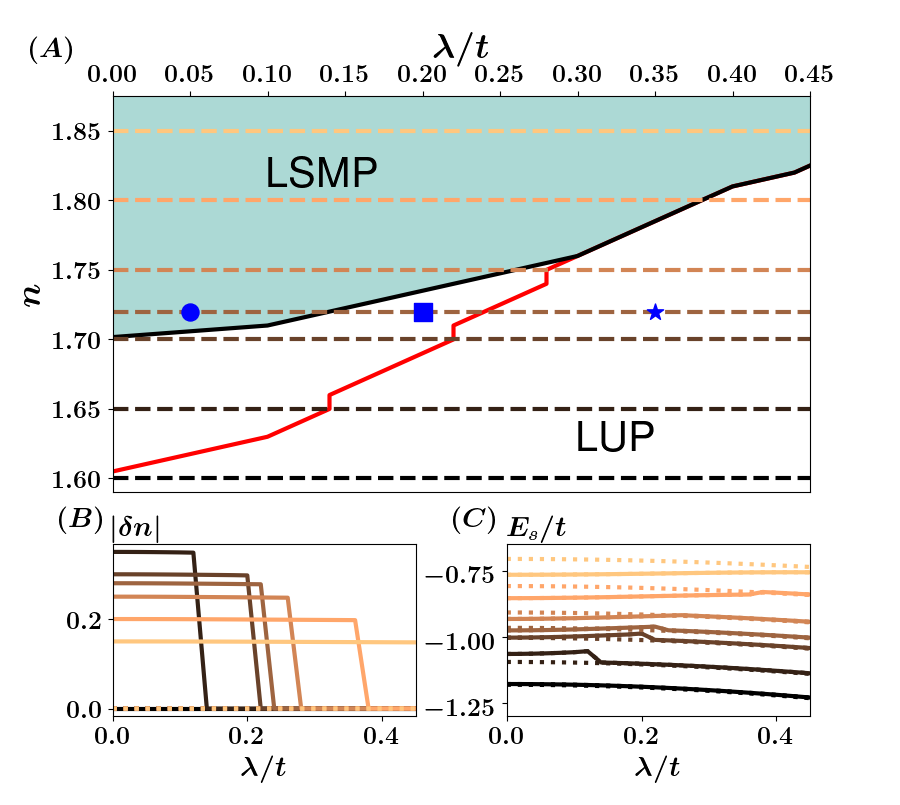}
\caption[0.5\textwidth]{(A) Displays the phase diagram as a function of electron density $n$ and inter-layer hopping $\lambda$ obtained for $U=20t$. Across all parameters, a layer uniform phase state exists, characterized by equal electron densities in both layers. However, a layer selective Mott phase emerges above the red line, with one layer spontaneously reaching half-filling. (B) Depicts the disparity in layer densities ($\delta n$) as a function of $\lambda$ for multiple average electron densities ($n$), with colors matching those in (A). For small $\lambda/t$, one layer closely approaches half-filling before dropping to equal density, indicating a sharp transition. (C) Compares the energy per site between the two solutions. The dashed lines are for the LUP solution, with the thick line corresponding to the LSMP one. The green area in (A) denotes where the LSMP has lower energy than the LUP solution. The region between black and red lines in (A) has a metastable LSMP phase.}
\label{fig:fig1}
\end{figure}
In Fig.~(\ref{fig:fig1}A), we present the phase diagram of the model in the total electron density $n$ and inter-layer hopping $\lambda$ plane, with $U=20t$ for each layer. The uniform phase exhibits identical layer electron densities, akin to the scenario described in ref.\cite{Haurie_2024} for disconnected layers. Such a uniform layer density solution is found across the entire phase diagram. However, a layer selective Mott phase emerges above the black line. In the LSMP phase, the density of one layer spontaneously approaches half-filling.

In Fig.~(\ref{fig:fig1}B) illustrates the difference in electron density between the two layers $\vert \delta n \vert$. Each line color corresponds to a specific total electron density associated with colored dashed lines in Fig.~(\ref{fig:fig1}A). Consequently, as $\lambda$ exceeds a critical value, $\delta n$ diminishes to zero, indicating a sharp transition to a state with equal density at each layer.
%At $\lambda=0$, the electron density at one of the layer is strictly at half-filling. We can gain insight into this solution by comparing it with the one-band Hubbard model. When one layer is half-filled, the chemical potential $\mu$ can be adjusted to ensure the self-consistent equations are met for the other layer. However, if $\mu$ evolved such that the half-filled band transitions to a metallic state, the self-consistent conditions for the previously half-filled layers are no longer satisfied, leading to the disappearance of the LSMP phase. When $\lambda\ne 0$, before transitioning to the layer uniform phase, one layer maintains a density close to half-filling for finite inter-layer hoppings. The critical $\lambda$ obtained from this analysis shows that the self-consistent LSMP solutions vanishes as the inter-layer hopping increases. 

However, the stability of the LSMP phase is determined by comparing the energy per site of the LSMP and the LUP phases. In Fig.\ref{fig:fig1}C), we show the energy per site for both states as a function of $\lambda$ across various electron densities, using the same color code as in Fig.(\ref{fig:fig1}B). The dashed line represents the LUP phase, while the continuous line depicts the LSMP. The LSMP phase becomes the ground state when its energy is lower than that of the uniform phase. The LSMP phase is energetically favorable near half-filling ($n=2$) and for weak to moderate inter-layer hopping.  The colored area of Fig.(\ref{fig:fig1}A) indicates the parameter range where the LSMP is the lower energy phase. Although the LSMP can stabilize below the thick black lines into the white region, it exhibits higher energy than the uniform phase. Only LUP solutions are observed below the thick red lines in Fig.~(\ref{fig:fig1}A). 

We can understand the origin of LSMP phase in the absence of inter-layer hopping i.e. $\lambda = 0$, where the two layers are disconnected. Note that the electron density in the two layers is constrained by the total electron density, i.e., $n=n_x + n_y$, which is determined by the same chemical potential, $\mu$. In the uniform phase, both layers share the same chemical potential which leads to the uniform solution of the single-layer Hubbard system with an electronic density of $n/2$~\cite{Haurie_2024}.

An alternative scenario arises when one layer becomes fully occupied spontaneously and falls entirely below the chemical potential for a given total electron density. To satisfy the density constraint, the chemical potential of the second layer adjusts to achieve a filling of $n_y = n - 1$. As the total electron density decreases, the chemical potential is reduced accordingly to adjust the filling of the metallic layer, although the band structure of the filled band remains unchanged. Once the chemical potential touches the top of the half-filled band, the system transitions to an uniform phase.

\begin{figure}[h!]
\includegraphics[width=8cm]{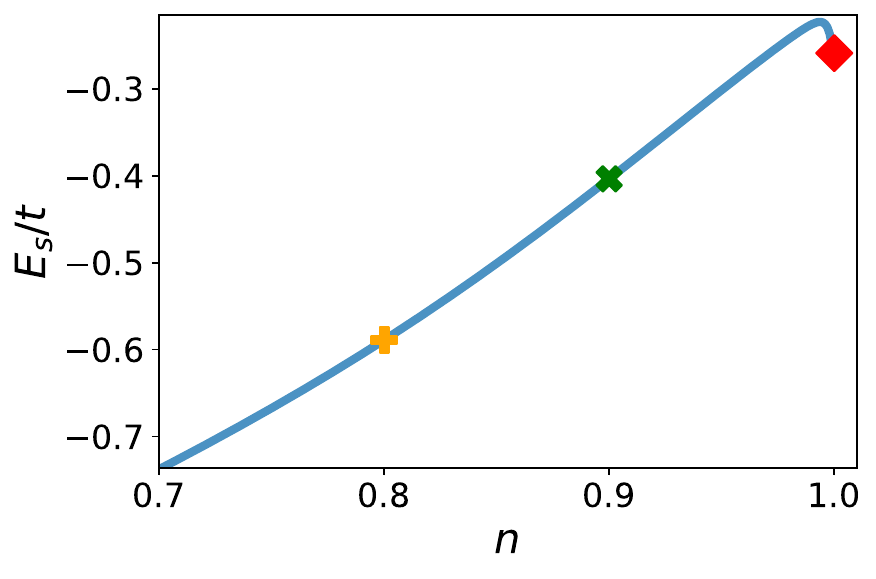}
\caption{Energy per site for the single-band Hubbard model as a function of electron density $n$. The red diamond, green square, and orange circle indicate the density and energy at $n=1.0$, $n=0.9$, and $n=0.8$ respectively. For two layers, if the the total electron density is $n=1.8$. Without interlayer hopping this demonstrates that it is energetically favorable to have one layer at $n=1$ and the other at $n=0.8$ (LSMP phase), rather than both layers at $n=0.9$ (OUP phase).}
\label{fig:EGS_n}
\end{figure}

Previous studies using the paramagnetic composite operator method for the single-band Hubbard model have observed negative compressibility near half-filling~\cite{Beenen0,Stanescu0} and a small energy dip near half-filling in other models~\cite{kagan2021electronic}, indicating that the uniform solution is unstable towards phase separation. In this work, phase separation occurs in the layer degrees of freedom, with one layer becoming half-filled while the other accommodates all the doped holes. Consequently, phase separation is the key mechanism driving the selective Mott phase in the paramagnetic limit. It remains unclear whether this observation is an artifact of the Hubbard operator method. Advancing the method to the next order, as suggested in Ref.~\cite{odashima2005high}, can provide insight into these questions.

The stability of the LSMP phase at $\lambda=0$ can be understood by studying the energy per site  $E_s$ of the one-band Hubbard model as a function of $n$. For $\lambda=0$, LSMP is favored phase whenever $E_{s}(1)+E_{s}(1-2\delta)<2E_{s}(1-\delta)$ with $\delta$ the doping. We checked that this is satisfied only close to half-filling indicating the regime of stability of the LSMP phase.

Consequently, we observe a spontaneous symmetry breaking of the layer degree of freedom in the bilayer Hubbard model. In the LSMP, one of the layers approaches the half-filled Mott state. As expected, such a phase is stabilized when the total electron density is close to half-filling~\footnote{Since at $n=2$, both the layers become Mott insulating.}. Additionally, the LSMP phase persists to a critical inter-layer hopping $\lambda_c$, indicating a sharp transition. Such critical $\lambda_c$ increases as we approach the particle-hole symmetric point. In Sec.~(\ref{Sec:withU}), we explore the dependence of the system for varying interaction strength $U$. Subsequent subsections characterize the two phases using multiple physical observables. 

In this section, we presented results for on-site inter-layer hopping. In App.~\ref{App:NN_hopping}, we demonstrate that the results remain qualitatively similar with different form of inter-layer hopping like extended $s$-wave and $d$-wave hopping.

\subsection{Bands and inter-layer hybridization}
\label{sec:Bands}

\begin{figure}[ht]
\includegraphics[width=0.5\textwidth]{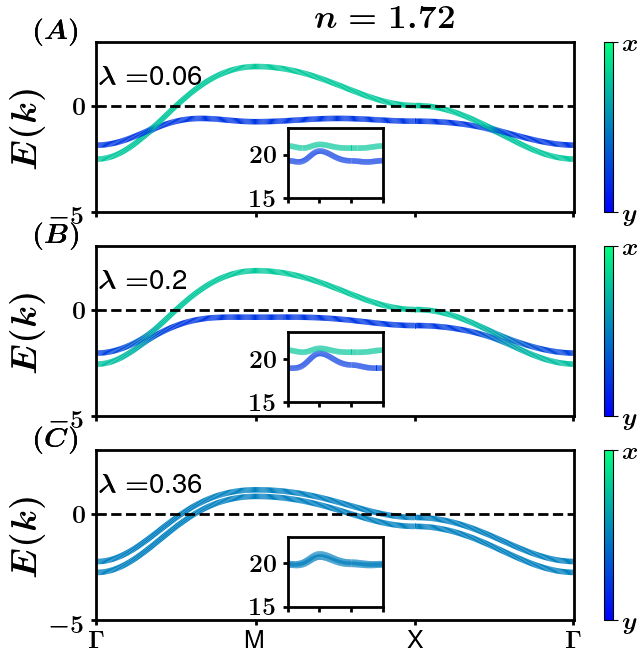}
\caption[0.5\textwidth]{Lower Hubbard bands at the high symmetry points for total electron density $n=1.72$ for (A) $\lambda=0.06t$ (B) $\lambda=0.2t$ (C) $\lambda=0.36t.$ These respectively corresponds to the blue circle, square and star from Fig.~(\ref{fig:fig1}). The colors indicate the layer character of the band at a particular momentum. In the LSMP phase, one band is below $E_F$ as expected and has mostly distinct feature of a single layer. Whereas in the uniform phase, both bands cross Fermi-energy and are completely hybridized. Inset shows the upper Hubbard bands.}
\label{fig:fig2}
\end{figure}

We present the lower Hubbard bands in the main panels of Fig.(\ref{fig:fig2}) for three different $\lambda$. In the LSMP phase, one of the bands flattens and remains below the Fermi energy $E_F$, while in the LUP phase, both bands cross $E_F$. The color of the lines shows the difference of $x$ and $y$ spectral function $\delta A=A_{xx}(\mathbf{k},\omega) -A_{yy}(\mathbf{k},\omega)$ thus characterizing the layer character of the bands if any. The bands are weakly hybridized at the intersection points for $\lambda=0.06t$  in Fig.~(\ref{fig:fig2}A). This shows that although the hybridization between the $x$ and $y$ layers vanishes at the Fermi level, it remains finite at lower energies. In Fig.~(\ref{fig:fig2}B), the hybridization increases as inter-layer hopping increases. Furthermore the Mott band approaches the Fermi-energy in Fig.~(\ref{fig:fig2}B) Finally, when we transition to a layer uniform phase, we lose the layer character of the electrons as shown in Fig.~(\ref{fig:fig2}C). We present the upper Hubbard bands as an inset of each panel, and they are separated from the lower Hubbard band by an energy gap of $\sim U$. Interestingly, the upper Hubbard bands remain entirely unhybridized in the LSMP phase.

\begin{figure}[ht]
\includegraphics[width=0.5\textwidth]{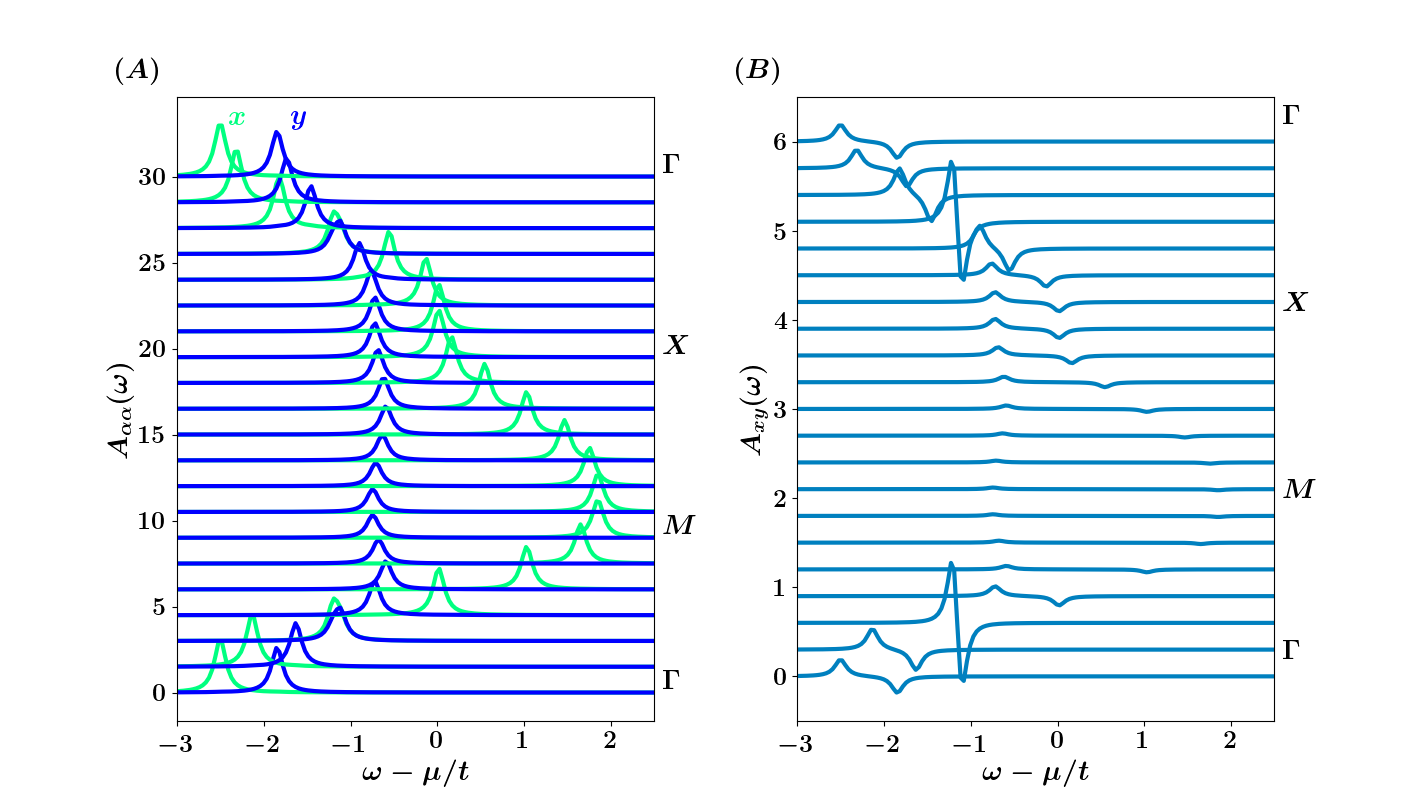}
\caption[0.5\textwidth]{Layer resolved spectral function of the lower Hubbard bands for $\lambda=0.06t$ and $n=1.72$, corresponding to the blue circle in the phase diagram in Fig.~(\ref{fig:fig1}A) displays the intra-layer spectral function $A_{\alpha \alpha}$ as a function of energy $\omega-\mu$ for different high symmetry $k$-points. The spectral for each $k$-points are shifted by $1.5t$ for clarity. (B) Shows the inter-layer spectral function $A_{\alpha \beta}$ for the same parameter. Here the results for each $k$-points are shifted by $0.3t$ for clarity.} 
\label{fig:fig3}
\end{figure}
To discern the nature of the layer and hybridization, we studied the spectral function in the LSMP phase. In Fig.~(\ref{fig:fig3}A), we show the intra-layer spectral functions for total electron density $n=1.72$ and $\lambda=0.06t$ for several high-symmetry k-points. The shift between each points is 1.5$t$ for $A_{\alpha \alpha}$ and $0.3t$ for $A_{\alpha\beta}$. The spectral function of the two layers overlaps for electrons near $\Gamma$-$X$ points of the Brillouin zone. The overlaps lead to a substantial inter-layer spectral function along the $\Gamma-X$ regions, as shown in Fig.~(\ref{fig:fig3}B). However, the spectral functions around $M-X$ points remain well separated and retain their layer character. Therefore, the inter-layer spectral function remains flat around $M-X$ regions.

Therefore, we reveal the characteristics of the bands in the LSMP and LUP phases. First, in the LSMP phase, one of the bands flattens and goes below the Fermi level. Secondly, in the LSMP phase, the two bands have weak hybridization and retain their layer character. Thirdly, the hybridization occurs only around specific momentum points where the two bands overlap. However, such hybridization remains appreciable for any finite $\lambda$, indicating the absence of a complete breakdown of hybridization in the LSMP phase.

\subsection{Fermi surface}
\label{sec:FS}
\begin{figure}[ht]
\includegraphics[width=0.499\textwidth]{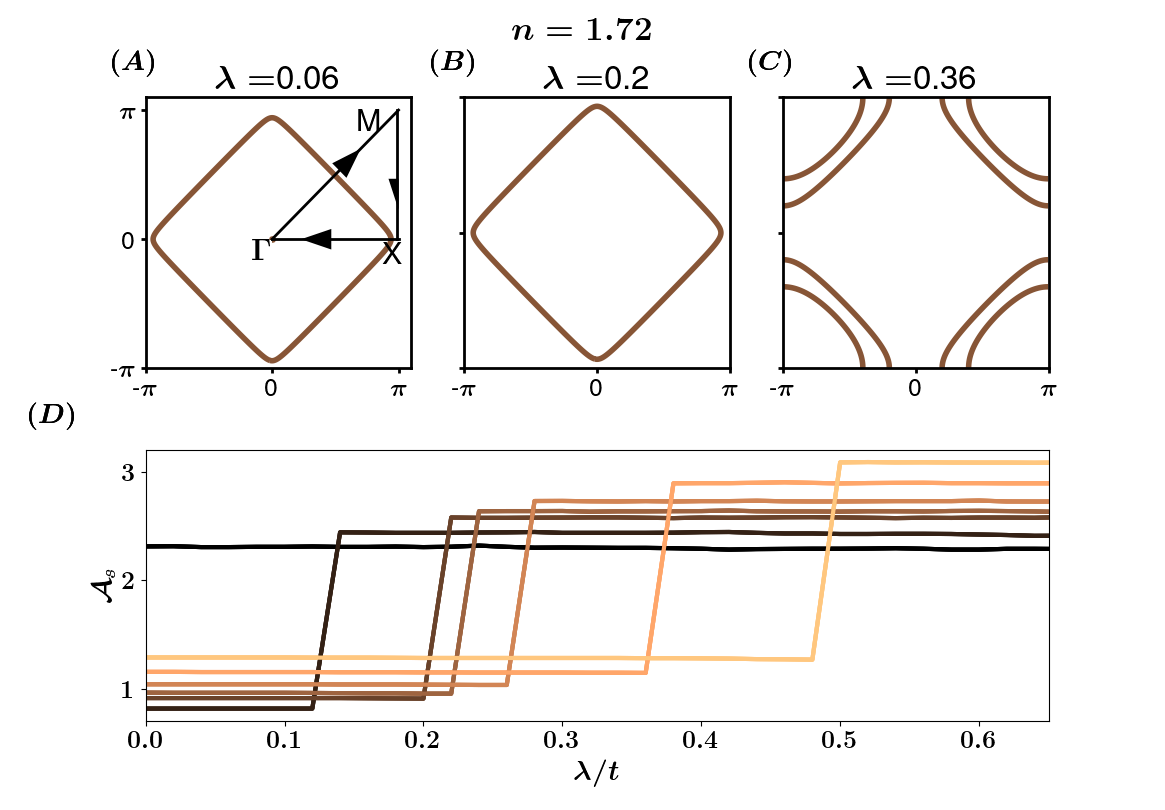}
\caption[0.5\textwidth]{(A) Displays the spectral function $A(\mathbf{k},E_F)$ at $\lambda=0.06t$ in the LSMP phase for $n=1.72$ (blue circle on Fig.~(\ref{fig:fig1}A) )\color{black}. (B) Depicts the same for $\lambda=0.2t$ (blue square on Fig.~( \ref{fig:fig1}A)), in the metastable region of the LSMP. (C) Fermi surface of the layer uniform phase at $\lambda=0.36t$ (blue star on Fig.~(\ref{fig:fig1}A)). In the LSMP, only a single Fermi sheet is observable, whereas in the metallic phase, two sheets are visible. Furthermore, the Fermi surface area is smaller in the LSMP phase compared to the LUP. (D) Illustrates the evolution of the Fermi volume $\mathcal{A}_s$ as a function of inter-layer hopping for different electron densities, showing a jump at the transition. In the LSMP phase, the Mott layer does not contribute to the Fermi volume since one band remains filled.}
\label{fig:fig4}
\end{figure}

In Fig.~(\ref{fig:fig4}A), we illustrate the Fermi surface contour, for $\lambda=0.06t$ at $n=1.72$ in the LSMP phase. In the LSMP phase, a single Fermi sheet is evident since the Mott layer remains fully occupied. With increasing inter-layer hopping there is little change in the Fermi surface as shown in Fig.~(\ref{fig:fig4}B). In contrast, both bands contribute to the Fermi surface in the metallic regime at $\lambda=0.36t$. Additionally, there is a significant increase in the size of the Fermi surface from the LSMP to the uniform phase. 

Moreover, we monitor the increase in the volume of the Fermi surface across the transition with inter-layer hopping for various total electron densities in Fig.~(\ref{fig:fig4}D). The volume exhibits a distinct jump at $\lambda_c$, which is more pronounced for lower $n$. From the evolution of the Fermi surface, the LSMP phase appears to be in a Kondo breakdown-like phase. However, we have shown in the previous section that hybridization of the two layers remains non-zero in the LSMP phase. 

A characteristic feature of non-Fermi liquid behavior is the violation of the Luttinger theorem, which asserts that the volume of the Fermi surface must equal the electron density. This theorem~\cite{oshikawa2000topological} applies to any system characterized by a low-energy theory of a Fermi liquid. As the composite excitations in our system are non-fermionic, a violation of the Luttinger theorem is expected~\cite{osborne2021broken,yang2022violation}. Consequently, our calculations demonstrate that the Luttinger theorem is violated in both the LSMP and strongly correlated LUP phase.

Hence, we observe a small Fermi surface in the LSMP to a large Fermi surface in the LUP phase. As a result, we detect a violation of the Luttinger theorem in the layer selective phase. Despite the sharp increase in the Fermi volume in the LUP phase, the Luttinger theorem continues to be violated. Thus, the traditional Fermi liquid picture is not applicable in the strongly correlated Hubbard model within this formalism.

\subsection{Density of states}
\label{sec:DOS}
\begin{figure}[ht]
\includegraphics[width=0.499\textwidth]{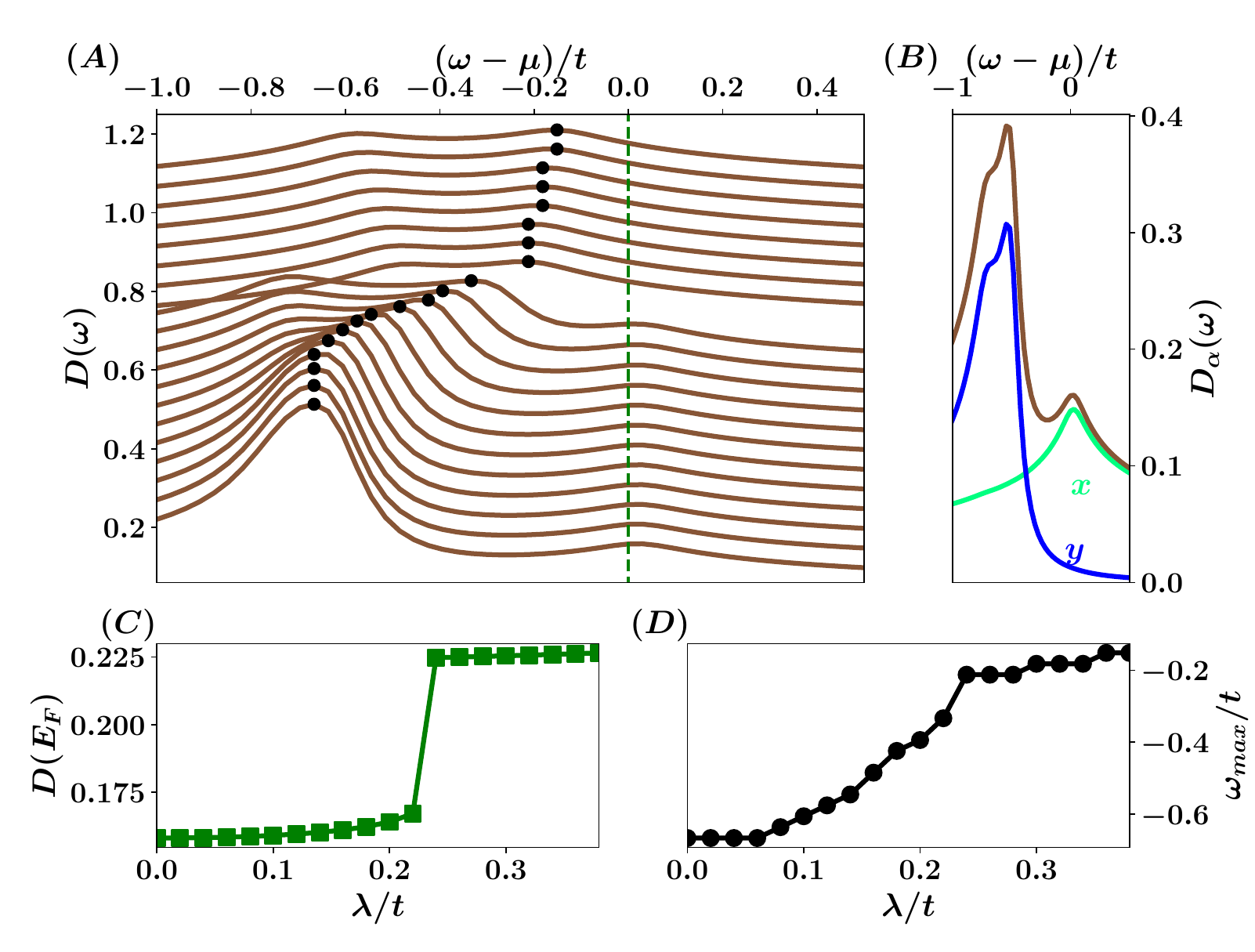}
\caption[0.5\textwidth]{(A) Presents the low-energy density of states at $n=1.72$ for several $\lambda$ ranging from $0$ to $0.42t$ with a $0.02t$ step. Each $\lambda$ curves are shifted vertically by $0.05t$ for clarity.$\lambda<0.25t$ corresponds to the LSMP phase. (B), shows the layer resolved density of states at $\lambda=0.16t$. Here, the $y$ layer approaches the Mott character, indicated by the reduced DOS at Fermi energy, whereas the $x$ layer remains metallic. The peak at the negative energy comes from the Mott layer. A progressive flattening of such peak of the LSMP occurs in (A) with increasing inter-layer hopping. (C) Shows the value of DOS at Fermi-energy, which manifest a jump around $\lambda=0.24t$, indicating the transition from LSMP to the uniform phase. (D) Tracks the energy associated with the Mott peak $\omega_{\rm max}$ (black dots in A). }
\label{fig:fig5}
\end{figure}
We now focus on the density of states (DOS). In Fig. (\ref{fig:fig5}A), we illustrate the low-energy density of states for $n=1.72$ across different $\lambda$ values. The curves for different $\lambda$ values are shifted vertically by $0.05t$ for clarity. Initially, at weak $\lambda$, the DOS features a prominent peak at negative energies (shown in black dots), with a finite DOS at the Fermi energy $E_F$. The presence of mobile electrons in one layer maintains an ungapped DOS at $E_F$. As inter-layer hopping increases, so does the DOS at Fermi energy. The transition from the LSMP to the LUP state is marked by a jump in the evolution of $D(E_F)$, as shown in Fig. (\ref{fig:fig5}C).

In Fig.~(\ref{fig:fig5}B), we depict the layer-resolved DOS for ${\lambda=0.17}$. The Mott layer $y$ displays a distinct peak at negative energy, accompanied by a small DOS at $E_F$. This peak signifies the energy required to extract electrons from the Mott layer. We track the evolution of the Mott peak $\omega_{\rm max}$ as a function of inter-layer hopping in Fig.~(\ref{fig:fig5}D). The energy $\omega_{\rm max}$ shows a smooth reduction. As a result of the hybridization between heavy Mott electrons and itinerant electrons, the peak flattens, making it challenging to discern the layer character of electrons prior to the transition to LUP.

Therefore, the LSMP phase features a reduced DOS at Fermi energy. The transition from the LSMP to the uniform phase is marked by a sharp increase in the states at the Fermi energy at a critical inter-layer hopping. Additionally, the LSMP phase displays a strong negative bias peak, which flattens, indicating the hybridization of Mott electrons with itinerant electrons. Notably, the flattening of the Mott peak with increasing inter-layer hopping occurs more smoothly than the sharp transitions observed in other quantities during the phase transition. This indicates that hybridization among the layers ultimately eliminates the layer differentiation.

\subsection{Quasiparticle residue}
\label{sec:QuasiRes}

\begin{figure}[h!]
\includegraphics[width=0.49\textwidth,height=0.45\textwidth]{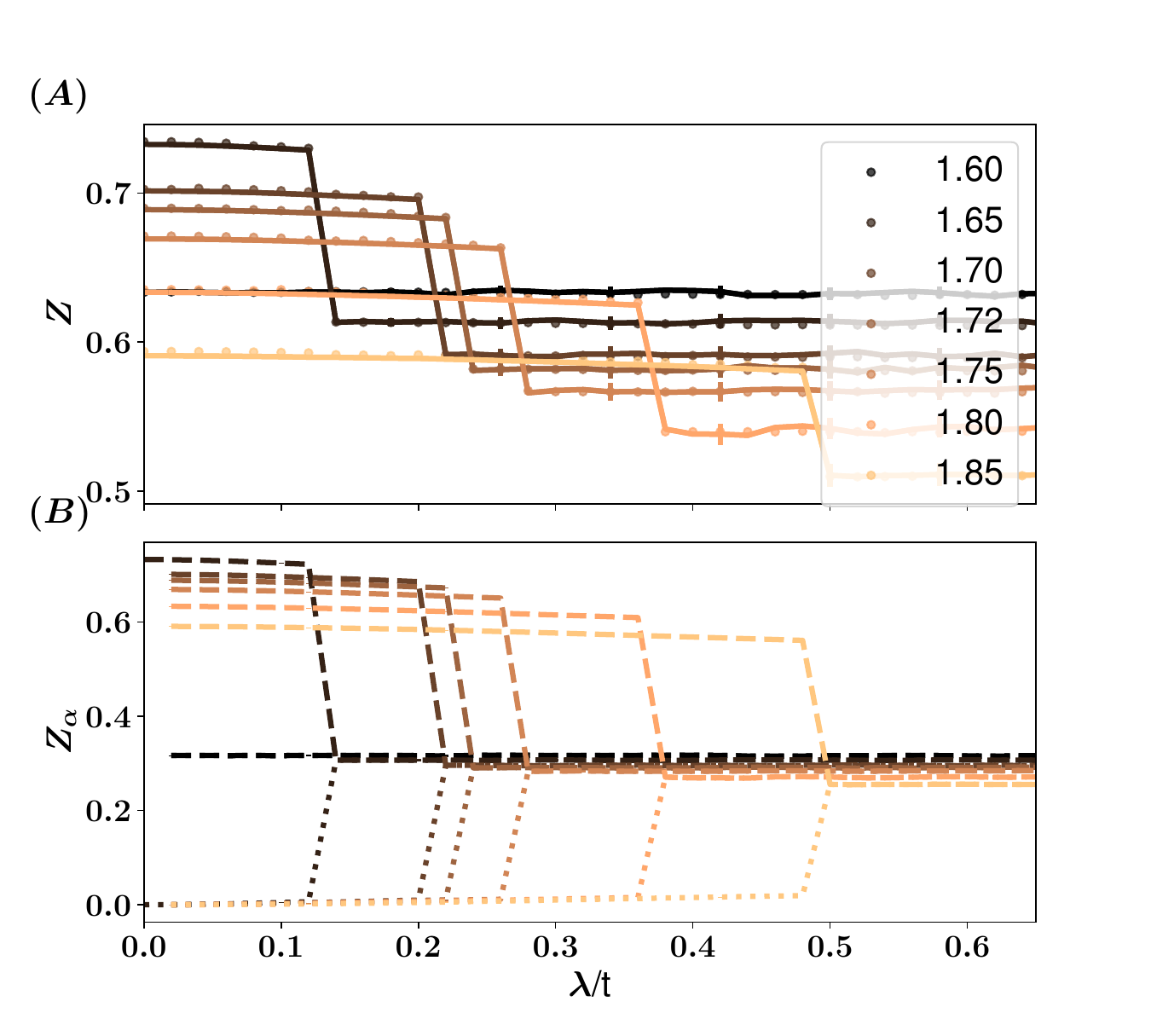}
\caption[0.5\textwidth]{(A)) Quasiparticle weight for several density as a function of $\lambda$, the errorbars are standard deviation when averaging over Fermi momentum. We overlapped with points the value of the jumps observable in $N_k$. (B) layer resolved quasiparticle weight. The Mott layer has a vanishing quasiparticle residue for finite $\lambda$.}
\label{fig:fig6}
\end{figure}
As the holon ($\xi$) and doublon ($\eta$) operators defined in Eq.~(\ref{eq:Xi_Operator}, \ref{eq:Eta_Operator}) are not satisfying fermionic anticommutation relations, the low energy theory in general is not a Fermi Liquid. This implies a violation of Luttinger theorem~\cite{Haurie_2024}. However we can compute the quasiparticle residue $Z$ by considering the residue of the Green function at the poles as detailed in Appendix~(\ref{Appendix:QW_Z}). Explicitly close to the Fermi surface, the Green function can be expanded in the following way
\begin{align}
&\mathcal{G}(\mathbf{k},\omega)\approx \dfrac{Z}{\omega+i\epsilon-\tilde{E}(\mathbf{k})+\dfrac{i}{2\tau(\mathbf{k},\omega)}}&
\end{align}
In Fig.~(\ref{fig:fig6}A), we present the total quasi-particle weight $Z=\sum \limits_n Z_{xx}^n + Z_{yy}^n$ as a function of $\lambda$ for several electron densities $n$. We calculated an average over the Fermi surface, with the displayed error bars corresponding to the standard deviation. Since $Z$ is never close to one, the quasiparticles of the theory are distinct from the bare electrons, highlighting the role of strong correlations. Eq.~(\ref{Eq:Qw}) indicates that the bare electrons are fractionalized into four quasiparticles associated with the bands.

We define the average electronic occupancy at a given momentum $k$, denoted as $N_k$, as follows:
\begin{align}
&N_k=\dfrac{1}{4}\sum\limits_{\alpha\sigma}\langle c^\dagger_{\alpha\sigma}c_{\alpha\sigma}\rangle&
\end{align}
We report the typical shape of $N_k$ along the $\Gamma$ to $M$ high symmetry line in Fig. (\ref{fig:figApp1}). By definition of the Green's function in Eq. \ref{Eq:Qw}, when crossing the Fermi contour a discontinuity of the order of $Z$ is observed in momentum space. One drop is observed for the LSMP, and the momentum resolved density does not vanish at $M$ due to the presence of the fully filled Mott band. Two drops are seen in the LUP due to the two Fermi contours of the phase. In this case, $Z$ is defined as the sum of these two drops. We report the $Z$ measured this way as a function of $\lambda$ on Fig (\ref{fig:fig6}A) as dots. There is a good agreement between the two ways of computing $Z$. Since there is no Fermi surface at half-filling, and that $Z$ is a quantity defined at the Fermi energy, $Z=0$ at half-filling. As electron density $n$ is decreasing, $Z$ increases. Therefore, we can also interpret the quasi-particle weight as a measurement of how close the system is from Mott regime, where interaction affect the system the most. 

Fig. (\ref{fig:fig6}B) displays the layer resolved quasi-particle weight. In the LSMP the Mott layer has an almost (exactly at $\lambda=0$) vanishing quasiparticle weight due to the localized nature of the electron. Since the other layer is further away from half-filling than the average doping its associated $Z$ increases from the average value. 

\subsection{Phase diagram with Hubbard interactions}
\label{Sec:withU}
\begin{figure}[h!]
    \centering
    \includegraphics[width=0.4\textwidth]{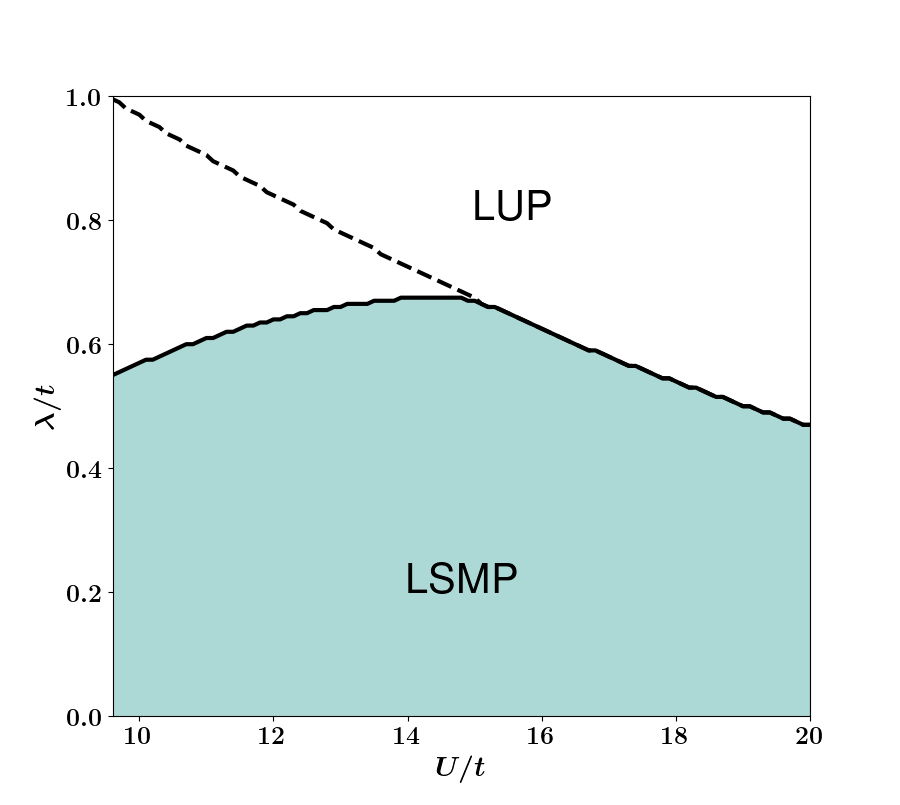}
    \caption{Here, we present the phase diagram as a function of intra-layer interaction strength $U$ and $\lambda$ for $n=1.85$. The colored area shows the region where the LSMP phase is energetically favorable, whereas the LUP phase is favorable in the white region. The LSMP is a metastable state between the thick and dashed lines.}
    \label{fig:fig7}
\end{figure}

Finally, In Fig.~(\ref{fig:fig7}), we present the phase diagram when the on-site repulsion $U$ is varied for electron density $n=1.85$. The colored region indicates where the LSMP phase is more favorable than the uniform phase. On the other hand, the LUP phase is stable in the white areas. The LSMP is a metastable state between the thick and dashed lines. 

For lower $U$, there is a weak increment of critical inter-layer hopping $\lambda_c$ where the LSMP phase becomes unfavorable. For large $U>15$ the same $\lambda_c$ reduces with increasing $U$. Thus, the critical $\lambda_c$ shows a nonmonotonic behavior with Hubbard interaction $U$. 

To qualitatively understand this result, we examine the destabilization of the LSMP phase. Initially, we assume double occupancy is forbidden due to strong repulsion. First, we focus on comprehending the decrease of $\lambda_c$ with increasing Hubbard interaction. 

In the LSMP phase, all the Mott layer sites are singly occupied, whereas the metallic layer has itinerant electrons and vacancies. When $\lambda \ne 0$, electrons can hop from the Mott to the metallic layers if the metallic site is vacant. A stable LSMP phase depends on refilling the holes in the Mott layer without significant hybridization. Without recombination, holes will proliferate in the Mott layer, leading to the breakdown of the layer differentiation. The hole created in the Mott layer is heavy and can be assumed to be approximately immobile.

Two scenarios are possible once an electron hops to the metallic layer: it either immediately hops back to the Mott layer or transfers to a neighboring site within the same layer. The stability of the LSMP phase depends on the ratio between $t$ and $\lambda$ for a fixed $U$ and electron density.

In the $\lambda \ll t$ regime, electrons are more likely to hop between neighboring sites of the same layers rather than between layers. Since double occupancy is forbidden, recombination can only happen when any electron on the metallic layer returns to the site at which Mott layer has the hole and goes back to the Mott layer. The mobility of electrons in the metallic layers is crucial in such a recombination pathway, which is relatively enhanced for lower $U$. Thus, the LSMP phase survives for a broader range of $\lambda$ when $U$ is lower. 

In the $\lambda \gg t$ regime, inter-layer hopping between the $x$ and $y$ layers dominates. This leads to strong hybridization between the $x$ and $y$ layers, which destroys the LSMP phase.

Finally, we focus on understanding why $\lambda_c$ decreases for $U < 15$. In this regime, double occupancy contributes to destabilizing the LSMP ordering. As $U$ decreases, double occupancies become more likely, increasing hybridization between the Mott and itinerant layers. Consequently, a metastable LSMP phase can still be present at lower $U$ due to the competing processes that are favored when lowering $U$. However, the LSMP becomes energetically less favorable compared to the uniform phase.

\section{Discussions\label{sec:Discussions}}
Studying the bilayer Hubbard model with strong on-site repulsion, we explored a spontaneous layer-selective phase in the vicinity of half-filling for weak to moderate inter-layer hopping. Despite the stability of a sizeable spontaneous layer symmetry broken phase for finite inter-layer hopping, the hybridization between the itinerant and localized layer remain nonvanishing. This finding aligns with results obtained using single-site DMFT for selective Mott phase~\cite{kugler2022orbital} and can be understood as the inevitability of electron transfer between the two layers(or orbitals), leading to finite hybridization. 

Therefore, our analysis demonstrates that selective Mott physics does not necessarily exhibit full Kondo breakdown transitions. This result does not contradict previous works that interpret Kondo breakdown as a selective Mott transition \cite{pepin2007kondo}. In that study \cite{pepin2007kondo}, the model involves free $c$-electrons interacting with strongly correlated $f$-electrons through inter-orbital hopping and inter-orbital Coulomb repulsion, leading to a complete breakdown phase. Although we do not find a complete Kondo breakdown phase within our model, hybridization is only finite where the two bands cross below the Fermi energy. Furthermore, significant layer differentiation and a small Fermi surface remain across a range of inter-layer hopping values. We confirmed the stability of our results against a non-zero inter-layer repulsion $U^\prime$ when it is a perturbation. While some proposals suggest the possibility of a complete Kondo breakdown for large $U^\prime$ \cite{pepin2008selective} is currently beyond the scope of this study. and remains an open question for the future. 

The Fermi volume jumps sharply at the transition between the LSMP and LUP phase. In LSMP, one of the bands spontaneously flattens and goes completely below the Fermi energy, thereby retaining the overall layer character~\cite{vojta2010orbital}. The presence of a filled band requires the other band to go below average density to maintain the total density. Consequently, a jump in the Fermi volume occurs when both bands are hybridized as the system transitions to layer uniform phase. Our finding indicates that any detection of a small Fermi surface in experiments is not equivalent to a complete breakdown of hybridization between two degrees of freedom. 

The idea of a critical Fermi surface~\cite{senthil2008critical} where the quasiparticle weight of one of the layers vanishes continuously does not apply to the abrupt LSMP transition observed in this study. A critical Fermi surface lacks well defined excitations due to the vanishing quasiparticle residue. Here, the transition is discontinuous, as evidenced by the sharp jumps in several quantities, and therefore, it lacks a quantum critical point. The entire Fermi surface abruptly disappears at the LSMP transition, and the quasiparticle residue of one layer vanishes at the transition.

This work highlights the possibility of stabilizing a spontaneous Mott selective phase within a bilayer system and only on-site Hubbard repulsion. Studies on selective Mott transitions focus on explicit symmetry breaking by tuning the crystal field splitting of the layers \cite{werner2007high,de2009orbital,jakobi2013orbital} or varying the intra-band Coulomb repulsion \cite{wu2008theory}. Other studies found spontaneous OSMP \cite{hoshino2017spontaneous} with a negative Hund's coupling $J_H$ in a three orbitals model. The LSMP transition is reminiscent of a discontinuous transition, with a large parameter region where the LSMP phase is metastable.

This study focuses on the paramagnetic solution of the bilayer Hubbard model. The localized layer in the LSMP, acting as quenched spins, are expected to order antiferromagnetically, similar to a single-band Hubbard model below the Néel temperature ~\cite{anderson1950antiferromagnetism,lee2006doping}. An intriguing future direction would be investigating this selective Mott phase in a lattice with frustrated spins, such as a triangular lattice, where magnetic order is suppressed. For an odd number of localized electrons per unit cell, the resulting paramagnetic spin liquid exhibits topological order~\cite{sachdev2008quantum}, featuring fractionalized spinon excitations coexisting with conduction electrons from the nonlocalized layer~\cite{vojta2010orbital}. Recently, many bilayer twisted materials have been shown to exhibit correlated behaviors analogous to Mott physics~\cite{po2018origin,cao2020tunable,wang2020correlated}. Extending the current study to twisted bilayer models is therefore a promising avenue to explore layer selective Mott phase in flat bands.

\section{Acknowledgement}
CP and AB thanks funding from CEFIPRA (Grant No. 6704-3). All computations are performed on the IPhT Kanta cluster. 
\appendix
\section{Roth decoupling}
\label{Appendix:Roth}
In this appendix we outline the Roth decoupling scheme \cite{roth1969electron} to compute two-point correlation functions like density-density, spin-spin and pair-hopping correlation function in Eq.~(\ref{eq:pab}) and hence $p_{\alpha\beta}$  in terms of $C^{0}_{ab}$ and $C^{1}_{ab}$. The Roth decoupling is an approximation scheme to compute the correlation for strongly correlated system unlike Wick's decomposition which works for weakly correlated systems. 

This scheme begins with the provided composite basis $\mathbf{\Psi}$. We call $\mathcal{B}_{i\alpha}=\{B^1_{i\alpha},B_{i\alpha}^2,\cdots,B_{i\alpha}^N\}$ the set of bilinear fermion operators at a site $i$ for the $\alpha$ layer and we suppress the spin index. For instance $B^1_{i\alpha}=c^\dagger_{i\alpha} c_{i\alpha}$ is the local density operator. Owing to the fermionic commutation relations,  the Roth decoupling scheme can be applied independently to $\langle B^{n_1}_{ix}B^{n_2}_{jx}\rangle$, $\langle B^{n_1}_{ix}B^{n_2}_{jy}\rangle$ and $\langle B^{n_1}_{iy}B^{n_2}_{jy}\rangle $ which means, in practice, there will be three independent sets of self-consistent equations for the $xx$, $xy$ and $yy$ sector. The solutions of these equations provide access to the 2-point correlation functions of each sector.

Roth decoupling allows us to find a set of self-consistent equations between the following static correlators $\langle B_{i\alpha}^n B_{j\sigma}^m\rangle$, $\forall n,m \in {1,\cdots,N}$ with $i$. Defining $\mathbf{c}_i=(c_{i\tau\uparrow},c_{i\tau\downarrow},c_{i\tau\uparrow}^\dagger,c_{i\tau\downarrow}^\dagger)$ the fermionic basis at each site $i$ for each layer $\tau$. then $\forall n$, we can express $B^n_{i\tau}$ as $B^n_{i\tau}=\mathbf{c}_{ia(n)}\mathbf{c}_{ib(n)}$ with two indices $a(n)$ and $b(n)$ that depend on $n$.

\vspace{1\baselineskip}

We choose $n,m,\alpha,\beta$ and decouple $\langle B_{i\alpha}^nB_{j\beta}^m\rangle$. This can be written as $B_{i\alpha}^n=\mathbf{c}_{ia(n)}\mathbf{c}_{ib(n)}$. 
Roth decoupling insight is to consider the following imaginary-time Green function $\mathds{D}_{\alpha jl}(\tau)$
\begin{align}
&\mathds{D}_{\alpha jl}(\tau)=-\left\langle T_\tau\left(\mathbf{\Psi}_{\alpha}(\tau)\mathbf{c}_{jb(n)}B^m_l\right) \right\rangle&
\end{align}
with $\alpha=(i,\delta)$ with $i$ a lattice sites and $\delta$ a components of the composite operator basis, such that the electronic operator $\mathbf{c}_{ia(n)}$ can be expressed as $\mathbf{c}_{ia(n)}=\mathbf{\Psi}_{(i,\delta_1)}+\mathbf{\Psi}_{(i,\delta_2)}$. The dependence of $\mathds{D}_{\alpha jl}(\tau)$ on $b(n)$ and $m$ is left implicit.  
The equation of motion for $\mathds{D}_{\alpha j l}$ is given by
\begin{align}
-\partial_\tau \mathds{D}_{\alpha j l}(\tau)&=\delta(\tau)\left\langle \left\{ \mathbf{\Psi}_\alpha(0),\mathbf{c}_{jb(n)}B_l^m\right\} \right\rangle \nonumber\\
&-\left\langle  T_\tau \left(\mathbf{j}(\tau)c_{j\sigma}^{(\dagger)}B_l^m\right)\right\rangle
\end{align}
Within the composite operator approximation $\mathbf{j}(\tau)=\mathds{E}\mathbf{\Psi}(\tau)$. We can define $\mathbf{f}_{\alpha jl}=\left\langle \left\{ \mathbf{\Psi}_\alpha(0),\mathbf{c}_{jb(n)}B_l^m\right\} \right\rangle$ which can be computed explicitly. 
\begin{align}
-\partial_\tau \mathds{D}_{\alpha j l}(\tau)=\delta(\tau)\mathbf{f}_{\alpha jl}+\sum\limits_\beta\mathds{E}_{\alpha\beta}\mathds{D}_{\beta jl}(\tau)
\end{align}
Next we can Fourier transform and after making analytic continuation $i\omega_n \rightarrow \omega+i\delta$ obtain

\begin{align}
&\sum\limits_{\beta}(\omega \delta_{\alpha\beta}-\mathds{E}_{\alpha\beta})\mathds{D}_{\beta jl}(\omega)=\mathbf{f}_{\alpha jl}&\\[5pt]
&\mathds{D}_{\alpha jl}(\omega)=\sum\limits_\beta \left(\omega \mathds{1}-\mathds{E}\right)^{-1}_{\alpha\beta}\mathbf{f}_{\beta jl}&
\end{align}
The relationship between $\mathds{E}$ and the composite Green function $\mathds{G}$ is $\mathds{G}_{\alpha\beta}(\omega)=\sum\limits_\gamma\left(\omega\mathds{1}-\mathds{E}\right)^{-1}_{\alpha\gamma}\mathds{I}_{\gamma\beta}$. In our approximation, $\mathds{I}$ is local. Expressing $\gamma=(q,\kappa)$ and $\beta=(h,\epsilon)$ with $q,h$ lattice sites and $\kappa,\epsilon$  components of the composite oeprator basis, locality implied $\mathds{I}_{\gamma\beta}=\delta_{qh}\hat{I}_{\kappa\epsilon}(i)$ , with $\hat{I}$ a $4\times 4$ matrix. 
%$\mathds{I}_{,j+mN}\propto \delta_{ij}$ with $n,m \in \{0,1\}$ and we define $\mathds{I}_{nm}(i)\equiv \mathds{I}_{i+nN,i+mN}$, $\mathds{I}_{mn}^{-1}(i)\equiv \mathds{I}^{-1}_{i+nN,i+mN}$

\begin{align}
\mathds{D}_{\alpha jl}(\omega)=\sum\limits_{\beta=(h,\epsilon) \kappa}\mathds{G}(\omega)_{\alpha  \beta} \hat{I}^{-1}(h)_{\epsilon \kappa}\mathbf{f}_{(h,\kappa) jl}
\end{align}

Applying the fluctuation-dissipation theorem, 
\begin{align}
\left\langle \mathbf{\Psi}_\alpha(0) c_{jb(n)}B_l^m\right\rangle=\int d\omega\left(1-n_F(\omega)\right)\dfrac{-1}{\pi}\text{Im}\big(\mathds{D}_{\alpha jl}(\omega)\big)
\end{align}
we derive a set of linear self-consistent equations for $\mathbf{f}_{\alpha jl}$ which constitutes the main result of the Roth decoupling scheme
\begin{align}
\mathbf{f}_{\alpha=(i,\delta) jl}=\sum\limits_{\beta=(h,\epsilon)\kappa}C^{\delta\epsilon}_{\mathbf{r}_i-\mathbf{r_j}}(i)\hat{I}^{-1}(h)_{\epsilon\kappa}\mathbf{f}_{(h,\kappa)jl}
\label{eq:f_Roth}
\end{align}

%If the three spatial indices $h,j,l$ are different $\mathbf{f}_{(h,\kappa),jl}=0$ due to the fermionic commutation relations. The sum over all the lattice sites in Eq.\ref{eq:f_Roth} thenreduces to a sum over only two values of  $h$, $h=j$ or $h=l$.

%\begin{align}
%\mathbf{f}_{i+nN,jl}=\sum\limits_{m,m'}&\left[C_{n+1,m+1}^{j-i}(i)\mathds{I}^{-1}(j)\mathbf{f}_{j+m'N,jl} \right. \nonumber \\
%& \left. +C_{n+1,m+1}^{l-i}(i)\mathds{I}^{-1}(l)\mathbf{f}_{l+m'N,jl}\right]
%\end{align}

To compute the initial correlation function $\langle B^n_{i\tau} B^m_{j\sigma}\rangle$, we then need to set $i=j$ and $l\rightarrow j$. 

%\begin{align}
%\mathbf{f}_{i+nN,ij}=\sum\limits_{m,m'}&\left[C_{n+1,m+1}^{0}(i)\mathds{I}^{-1}(i)\mathbf{f}_{i+m'N,ij} \right. \nonumber \\ &\left.+C_{n+1,m+1}^{j-i}(i)\mathds{I}^{-1}(j)\mathbf{f}_{j+m'N,ij}\right]
%\label{eq:Final_f}
%\end{align}

\begin{align}
\mathbf{f}_{\alpha=(i,\delta) ij}=\sum\limits_{\beta=(h,\epsilon)\kappa}C^{\delta\epsilon}_{\mathbf{r}_i-\mathbf{r_j}}(i)\hat{I}^{-1}(h)_{\epsilon\kappa}\mathbf{f}_{(h,\kappa)ij}
\label{eq:f_Roth_final}
\end{align}

We can introduce the two-particle static correlation function vector in each layer sector
\begin{align}
\mathbf{G}_{\alpha\beta}=\left(\langle \Delta_{i\alpha}\Delta_{j\beta}^\dagger\rangle, \langle  S_{i\alpha}^-S_{j\beta}^+\rangle, \langle n_{i\alpha\uparrow}n_{j\beta\uparrow}\rangle, \langle n_{i\alpha\uparrow}n_{j\beta\downarrow}\rangle\right)^T
\end{align}
Then the Roth self-consistent equations~\ref{eq:f_Roth_final} can be arranged in the following form
\begin{align}
\label{eq:Roth}
&\mathds{A}_{\alpha\beta}\mathbf{G}_{\alpha\beta}=\mathbf{B}_{\alpha\beta}&
\end{align}
with $\mathds{A}_{\alpha\beta}$ a $4\times 4$ matrix acting in the static correlation function space and $\mathbf{B}_{\alpha\beta}$ a size $4$ vector. Eq.~\ref{eq:Roth} can be inverted to obtain an explicit expression of the relevant static correlation function to compute $p_{\alpha\beta}$. Note that it is given by
\begin{align}
&p_{\alpha  \beta}=\frac{1}{4N}\sum_{\langle i,j \rangle}\langle \hat{n}_{i \alpha \sigma}\hat{n}_{j \beta \sigma}\rangle+\langle S^-_{i \alpha }S^+_{j \beta}\rangle-\langle \Delta_{i \alpha }\Delta^\dagger_{j \beta}\rangle&
\end{align}
The individual two-point expectation values can be written as follow
\begin{align}
&\langle n_{\alpha\sigma}n_{\beta\sigma} \rangle=\dfrac{\Big(-1-\phi_{\alpha\beta}+2\big(\rho_{0,\alpha\beta}+\phi_{\alpha\beta}(\rho_{0,\alpha\beta})+\rho^{n_\sigma n_\sigma} _{\alpha\beta}\big)\Big)}{-2+2 (\phi_{\alpha\beta})^2} \label{eq:nn}&\\
&\langle S^-_{\alpha}S^+_{\beta} \rangle=-\dfrac{\rho^S_{\alpha\beta}}{1+\phi_{\alpha\beta}}\label{eq:ss}&\\[5pt]
&\langle\Delta_{\alpha}\Delta^\dagger_{\beta} \rangle=\dfrac{\rho^\Delta_{\alpha\beta}}{1-\phi_{\alpha\beta}}\label{eq:dd}&
\end{align}
where we have defined
\begin{align}
&\phi_{xx}=\phi_{xy}^\Delta=\dfrac{2}{2-n_x}\big(C_0^{11}+C_0^{21}\big)-\dfrac{2}{n_x}\big(C_0^{12}+C_0^{22}\big)&\\[5pt]
&\phi_{yy}=\phi_{yx}^\Delta=\dfrac{2}{2-n_y}\big(C_0^{33}+C_0^{43}\big)-\dfrac{2}{n_y}\big(C_0^{34}+C_0^{44}\big)&\\[5pt]
&\rho_{0,xx}=\dfrac{2}{2-n_x}\dfrac{n_x}{2}\big(C_0^{11}+C_0^{21}\big)&\\[5pt]
&\rho_{0,yy}=\dfrac{2}{2-n_y}\dfrac{n_y}{2}\big(C_0^{33}+C_0^{43}\big)&\\[5pt]
&\rho_{0,xy}=\dfrac{2}{2-n_x}\dfrac{n_y}{2}\big(C_0^{11}+C_0^{21}\big)&\\[5pt]
&\rho_{0,yx}=\dfrac{2}{2-n_y}\dfrac{n_x}{2}\big(C_0^{33}+C_0^{43}\big)&
\end{align}

\begin{align}
&\rho_{xx}^\Delta=\dfrac{2}{2-n_x}\big(C_1^{11}+C_1^{21}\big)\big(C_1^{21}+C_1^{22}\big)&\\[2pt]
\nonumber&+\dfrac{2}{n_x}\big(C_1^{11}+C_1^{12}\big)\big(C_1^{12}+C_1^{22}\big)&\\[5pt]
&\rho_{yy}^\Delta=\dfrac{2}{2-n_y}\big(C_1^{33}+C_1^{43}\big)\big(C_1^{43}+C_1^{44}\big)&\\[2pt]
\nonumber&+\dfrac{2}{n_y}\big(C_1^{33}+C_1^{34}\big)\big(C_1^{34}+C_1^{44}\big)&\\[5pt]
&\rho_{xy}^\Delta=\dfrac{2}{2-n_y}\big(C_0^{13}+C_0^{23}\big)\big(C_0^{41}+C_0^{42}\big)&\\[2pt]
\nonumber&+\dfrac{2}{n_y}\big(C_0^{31}+C_0^{32}\big)\big(C_0^{14}+C_0^{24}\big)&\\[5pt]
&\rho_{yx}^\Delta=\dfrac{2}{2-n_x}\big(C_0^{31}+C_0^{41}\big)\big(C_0^{23}+C_0^{24}\big)&\\[2pt]
\nonumber&+\dfrac{2}{n_x}\big(C_0^{13}+C_0^{14}\big)\big(C_0^{32}+C_0^{42}\big)&
\end{align}
\begin{align}
&\rho_{xx}^S=\dfrac{2}{2-n_x}\big(C_1^{11}+C_1^{21}\big)\big(C_1^{11}+C_1^{12}\big)&\\[2pt]
\nonumber&+\dfrac{2}{n_x}\big(C_1^{22}+C_1^{12}\big)\big(C_1^{21}+C_1^{22}\big)&\\[5pt]
&\rho_{yy}^S=\dfrac{2}{2-n_y}\big(C_1^{33}+C_1^{43}\big)\big(C_1^{33}+C_1^{34}\big)&\\[2pt]
\nonumber&+\dfrac{2}{n_y}\big(C_1^{44}+C_1^{34}\big)\big(C_1^{43}+C_1^{44}\big)&\\[5pt]
&\rho_{xy}^S=\dfrac{2}{2-n_y}\big(C_0^{23}+C_0^{13}\big)\big(C_0^{31}+C_0^{32}\big)&\\[2pt]
\nonumber&+\dfrac{2}{n_y}\big(C_0^{14}+C_0^{24}\big)\big(C_0^{41}+C_0^{42}\big)&\\[5pt]
&\rho_{yx}^S=\dfrac{2}{2-n_x}\big(C_0^{41}+C_0^{31}\big)\big(C_0^{13}+C_0^{14}\big)&\\[2pt]
\nonumber&+\dfrac{2}{n_x}\big(C_0^{32}+C_0^{42}\big)\big(C_0^{23}+C_0^{24}\big)&
\end{align}
\begin{align}
&\rho_{xx}^{n_\sigma n_\sigma}=\dfrac{2}{2-n_x}\big(C_1^{11}+C_1^{21}\big)\big(C_1^{11}+C_1^{12}\big)&\\[2pt]
\nonumber&+\dfrac{2}{n_x}\big(C_1^{22}+C_1^{12}\big)\big(C_1^{21}+C_1^{22}\big)&\\[5pt]
&\rho_{yy}^{n_\sigma n_\sigma}=\dfrac{2}{2-n_y}\big(C_1^{33}+C_1^{43}\big)\big(C_1^{33}+C_1^{34}\big)&\\[2pt]
\nonumber&+\dfrac{2}{n_y}\big(C_1^{44}+C_1^{34}\big)\big(C_1^{43}+C_1^{44}\big)&\\[5pt]
&\rho_{xy}^{n_\sigma n_\sigma}=\dfrac{2}{2-n_y}\big(C_0^{13}+C_0^{23}\big)\big(C_0^{31}+C_0^{32}\big)&\\[2pt]
\nonumber&+\dfrac{2}{n_y}\big(C_0^{14}+C_0^{24}\big)\big(C_0^{41}+C_0^{42}\big)&\\[5pt]
&\rho_{yx}^{n_\sigma n_\sigma}=\dfrac{2}{2-n_x}\big(C_0^{31}+C_0^{41}\big)\big(C_0^{13}+C_0^{14}\big)&\\[2pt]
\nonumber&+\dfrac{2}{n_x}\big(C_0^{32}+C_0^{42}\big)\big(C_0^{23}+C_0^{24}\big)
\end{align}

\section{Physical quantities within composite operator method\label{Appendix:Physical_Quantities}}
\subsection{Energy per-site}
\label{Appendix:GSE}
The expectation value of the Hamiltonian gives the energy.
\begin{align}
E_s=&-t\sum\limits_{\langle ij\rangle\alpha\sigma}\langle c^\dagger_{i\alpha\sigma}c_{j\alpha\sigma}+h.c. \rangle -\lambda \sum\limits_{i\sigma}\langle c^\dagger_{ix\sigma}c_{iy\sigma}+h.c.\rangle& \nonumber \\
&+U\sum\limits_{i\alpha\sigma}\langle\hat{n}_{i\alpha\uparrow}\hat{n}_{i\alpha\downarrow}\rangle
\end{align}
The first term is the intralayer kinetic energy $\langle \mathcal{H}_t \rangle$ and the second term  is inter-layer kinetic energy $\langle \mathcal{H}_{\lambda} \rangle$. The third term is the potential energy $\langle \mathcal{H}_U \rangle$. We evaluate in terms of the correlation function
\begin{align}
    \langle \mathcal{H}_t \rangle &= 8t  \sum_{i,j=1}^2  \left( C_1^{ij} + C_1^{(2i)(2j)} \right) \\
    \langle \mathcal{H}_\lambda \rangle &= 4\lambda \left(C_0^{13}+C_0^{23}+C_0^{14}+C_0^{24} \right) \\
    \langle \mathcal{H}_U \rangle &= U \left(\frac{n}{2}-C_0^{21}-C_0^{43} - C_0^{22} - C_0^{44} \right)
\end{align}

\subsection{Spectral function and density of states}
\label{Appendix:LDOS}
The electronic Green's function can be calculated from the composite Green's function  $\mathds{G}(\omega)$ using the relation $c_{i\alpha\sigma}=\eta_{i\alpha\sigma}+\xi_{i\alpha\sigma}$. Therefore, 
\begin{align}
&\mathcal{G}_{\alpha\beta}(\mathbf{k},\omega)=\sum\limits_{\psi(\alpha)\psi(\beta)}\mathds{G}_{\psi(\alpha)\psi'(\beta)}(\mathbf{k},\omega)&
\end{align}
with $\psi(\alpha),\psi'(\beta)$ representing a summation over the composite operator of layer character $\alpha\beta$.
The spectral function is given by
\begin{align}
    A_{\alpha\beta}(\mathbf{k},\omega)=-\frac{1}{\pi} \text{ Im}\big[ \mathcal{G}_{\alpha\beta}(\mathbf{k},\omega)\big]
\end{align}
The total intralayer spectral function is given by ${A(\mathbf{k},\omega)=\sum_\alpha A_{\alpha \alpha}(\mathbf{k},\omega)}$. Similarly the inter-layer spectral function is given by ${A_\delta(\mathbf{k},\omega)=\sum_\alpha A_{\alpha \bar{\alpha}}(\mathbf{k},\omega)}$.
The layer resolved density of states can be evaluated as,
\begin{align}
&D_{\alpha}(\omega)=-\frac{1}{\pi} \int d\mathbf{k} \text{ Im}\big[ \mathcal{G}_{\alpha\alpha}(\mathbf{k},\omega)\big]&
\end{align}

\begin{figure}[ht]
\includegraphics[width=0.499\textwidth]{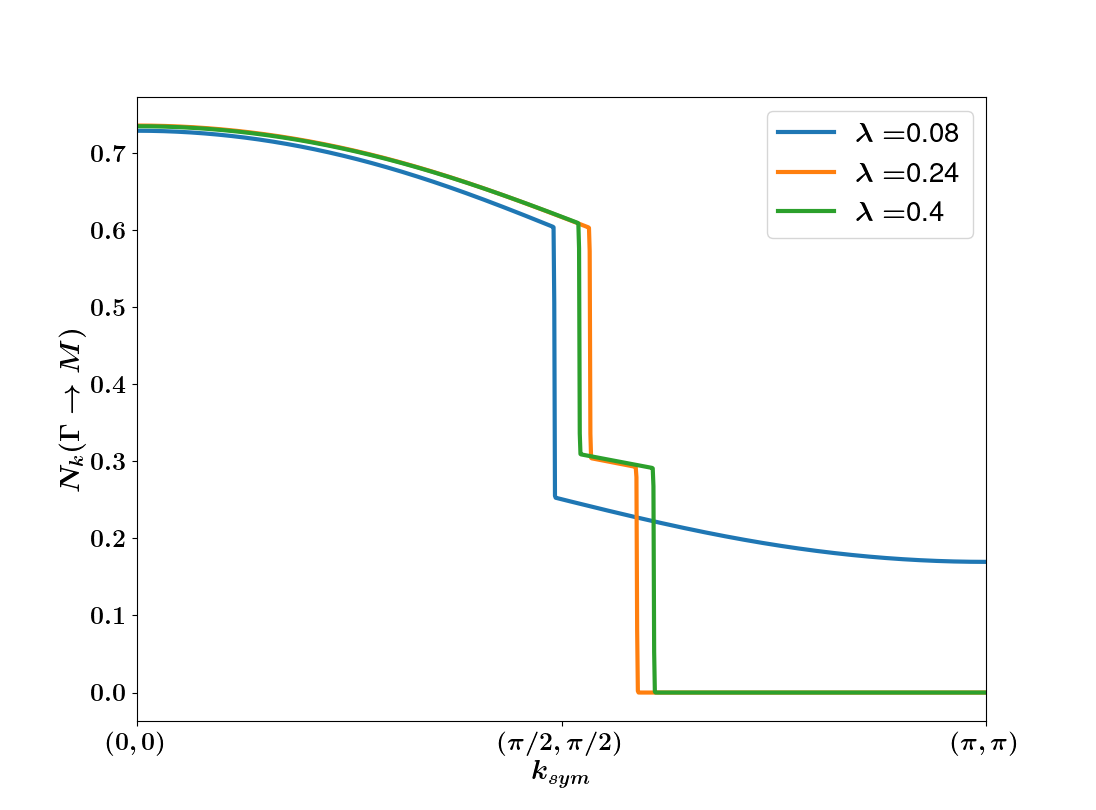}
\caption[0.5\textwidth]{Momentum resolved electron density along the $\Gamma$ to M high symmetry line. The electron density is fixed at $n=1.7$ and the blue curve corresponds to the LSMP while the two other are in the 'Fermi liquid' phase. In the vicinity of Fermi momentum a jump is observable. Since the LUP phase exhibits two electron pockets, two distinct jumps are observable. The LSMP only presents one jump, but the density does not vanish at M due to the fully filled layer. According to Fermi liquid theory the values of these jumps are the quasiparticle weights. On \ref{fig:fig5} (A) we report the value of the jump as dots, and confirm it matches the other definition of the quasiparticle weight.}
\label{fig:figApp1}
\end{figure}

\subsection{QuasiParticle-Weight $Z$\label{Appendix:QW_Z}}
The single particle Green function can in general be written
\begin{align}
&\mathcal{G}(\mathbf{k},\omega)=\dfrac{1}{\omega-E(\mathbf{k})-\Sigma(\mathbf{k},\omega)}&
\end{align}
with $\Sigma(\mathbf{k},\omega)$ the electronic self-energy.
For $\mathbf{k}$ close to the Fermi surface and at small energies, the electronic Green's function can be expanded as
\begin{align}
&\mathcal{G}(\mathbf{k},\omega)\approx \dfrac{Z}{\omega-\tilde{E}(\mathbf{k})+\dfrac{i}{2\tau(\mathbf{k},\omega)}}&
\end{align}
with $Z$ the quasiparticle weight, $\tilde{E}(\mathbf{k})$ the renormalized quasiparticle energy and $\tau(\mathbf{k},\omega)$ the quasiparticle lifetime which should diverge as $\omega\rightarrow 0$ for a Fermi Liquid. The quasiparticle-weight $Z$ can be computed in terms of the self energy as
\begin{align}
&Z^{-1}=1-\dfrac{\partial}{\partial\omega}\text{Re}\big[\Sigma(k_F,\omega)\big]|_{\omega=0}&
\end{align}
In the COM formalism, $Z$ can be computed analytically. The COM Green function read
\begin{align}
&\mathds{G}_{\gamma\nu}(\mathbf{k},\omega)=\sum\limits_{n=1}^4\dfrac{\sigma^n_{\gamma\nu}(\mathbf{k})}{\omega-E_n(\mathbf{k})+i\delta}&
\end{align}
From $\mathds{G}$, we can compute $\mathcal{G}$ as 

\begin{align}
&\mathcal{G}_{\alpha\beta}(\mathbf{k},\omega)=\sum\limits_{n=1}^4\dfrac{1}{\omega-E_n(\mathbf{k})+i\delta}\sum\limits_{\psi(\alpha)\psi'(\beta)}\sigma_{\psi(\alpha)\psi(\beta)}^n(\mathbf{k})&
\end{align}
with $\psi(\alpha),\psi'(\beta)$ representing a summation over the composite operator of layer character $\alpha\beta$. This can be rewritten as

\begin{align}
&\mathcal{G}(\mathbf{k},\omega)=\sum\limits_{n=1}^4\dfrac{Z^n(\mathbf{k})}{\omega-E_n(\mathbf{k})+i\delta}&\\[5pt]
&Z_{\alpha\beta}^{n_0}(\mathbf{k})=\sum\limits_{\psi(\alpha)\psi(\beta)}\sigma^{n_0}_{\psi(\alpha)\psi(\beta)}(\mathbf{k})&
\label{Eq:Qw}
\end{align}
This expression for the Green function clearly demonstrates the fractionalization of electrons with this method, with the electronic weight distributed over two bands separated by a gap of order $U$ the Coulomb repulsion.
In the limit $\omega \rightarrow E_n(\mathbf{k})$, $\dfrac{Z^n(\mathbf{k})}{\omega-E_n(\mathbf{k})+i\delta} \gg \dfrac{Z^{n_0}(\mathbf{k})}{\omega-E_{n_0}(\mathbf{k})+i\delta}$ with $n\ne n_0$, and we can conclude that the quasiparticle-weight at this $k_F$ is given by
\begin{align}
&Z_{\alpha\beta}=\sum\limits_{\psi(\alpha)\psi(\beta)}\sigma^{n_0}_{\psi(\alpha)\psi(\beta)}(k_F)&
\end{align}
This weight has to notable property according to Fermi liquid theory of being equal to the gap observed in momentum resolved electron density $N_k$ when crossing the Fermi surface. 

In Fig. \ref{fig:figApp1}, $N_k$ is plotted along the $\Gamma$ to M line for several $\lambda$ at $n=1.70$. 

\section{LSMP for different inter-layer hopping symmetry\label{App:NN_hopping}}
For systems with different interlayer orbital symmetry direct interlayer-site hopping is usually forbidden due to symmetry constraints between different orbitals. As a result, inter-layer nearest-neighbor hopping becomes the dominant term, which can destabilize the layer-selective Mott phase~\cite{georges2013strong}. In this appendix, we examine both extended $s$- and $d$-wave inter-layer hopping and demonstrate that it produces similar physical behavior as on-site inter-layer hopping.
The extended $s$- and $d$-wave interlayer hopping read respectively in moemtum space as
\begin{align}
&\mathcal{H}_\lambda^s(\mathbf{k})=-2\lambda\sum \limits_{\mathbf{k}\sigma} \left(\cos(k_x)+\cos(k_y)\right)\left(c^\dagger_{\mathbf{k}x\sigma}c_{\mathbf{k}y\sigma}+ h.c. \right)&\\
&\mathcal{H}_\lambda^d(\mathbf{k})=-2\lambda\sum \limits_{\mathbf{k}\sigma} \left(\cos(k_x)-\cos(k_y)\right)\left(c^\dagger_{\mathbf{k}x\sigma}c_{\mathbf{k}y\sigma}+ h.c.\right)&
\end{align}
\begin{figure}[ht]
\includegraphics[width=0.499\textwidth]{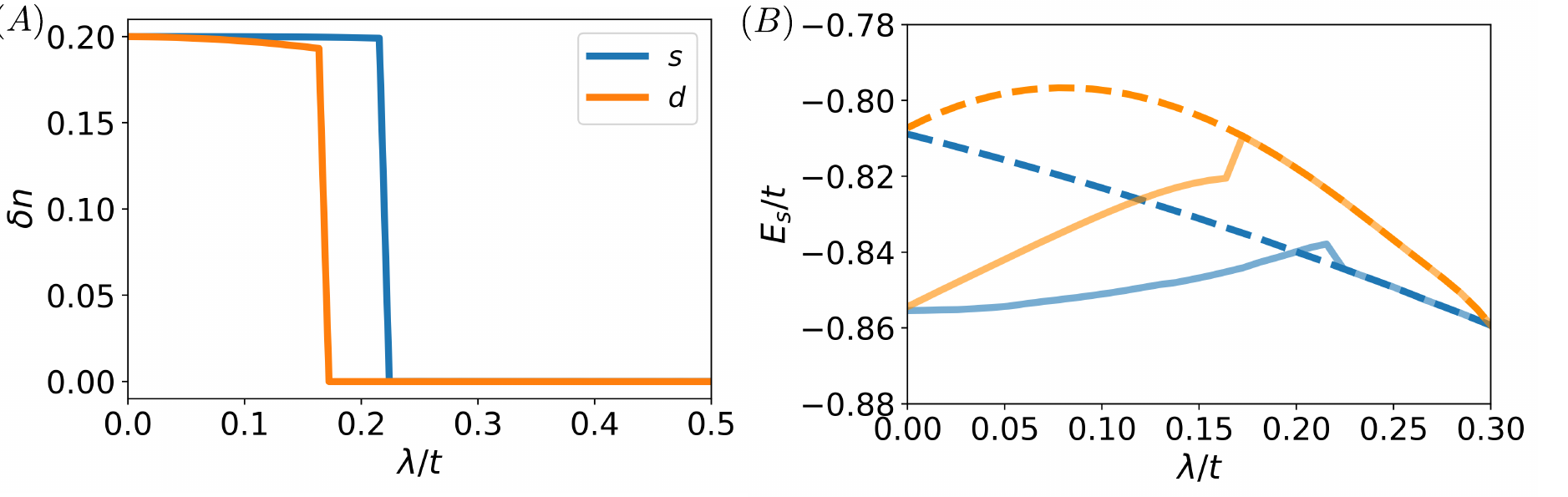}
\caption{(A) Disparity in layer densities ($\delta n$) as a function of $\lambda$ at $n=1.8$ and $U=20t$ for both extended $s$-wave (blue curve) and $d$-wave (orange curve) nearest-neighbor hopping. (B) Compares the energy per site between the LUP and LSMP phase for both $s$-wave (blue curve) and $d$-wave (orange curve) nearest-neighbor hopping. Dashed lines are for the LUP solutions and thick lines for the LSMP ones.} 
\label{fig:deltan_NN}
\end{figure}
In Fig.~(\ref{fig:deltan_NN}a), we show that for a fixed $U=20t$, we observe a transition from the LSMP to LUP phase at low doping. Similar to the direct on-site interlayer hopping the LSMP is energetically favorable for small $\lambda$ as shown in Fig.~(\ref{fig:deltan_NN} b) for both different form factor of the nearest neighbor hopping.

%\bibliography{Refs_OSMP.bib}


\end{document}